\documentclass[pra,twocolumn,preprintnumbers,amsmath,amssymb,superscriptaddress,showpacs,longbibliography]{revtex4-2}
\usepackage{graphicx}
\usepackage{amsmath}
\usepackage{graphics}
\usepackage{amssymb}
\usepackage{amsfonts,epsfig}
\usepackage{dsfont}
\usepackage{natbib}
\usepackage{hyperref}
\usepackage{enumitem}
\usepackage{physics}
\usepackage{color}
\usepackage{relsize}
\usepackage{bbm}
\usepackage[mathscr]{eucal}

\allowdisplaybreaks

\newcommand{\beq}{\begin{equation}}
	
	\newcommand{\eeq}{\end{equation}}
\newcommand{\bea}{\begin{eqnarray}}
	\newcommand{\eea}{\end{eqnarray}}

\newcommand{\HH}{\hat{H}}
\newcommand{\Sz}{\hat{S}_z}
\newcommand{\I}{\hat{I}}

\newcommand{\densmat}{\hat{\rho}}
\usepackage{color}

\newcommand{\U}{\hat{\mathcal{U}}}
\usepackage{ulem} 

\newcommand{\F}{\mathcal{F}}
\renewcommand{\I}{\hat{I}}

\begin{document}
	
	\title{Appearance of objectivity for NV centers interacting with dynamically polarized nuclear environment}
	\author{Damian Kwiatkowski}
	\affiliation{Institute of Physics, Polish Academy of Sciences, al.~Lotnik{\'o}w 32/46, PL 02-668 Warsaw, Poland}
	\affiliation{QuTech, Delft University of Technology, 2628 CJ Delft, Netherlands}
  \affiliation{Kavli Institute of Nanoscience, Delft University of Technology, 2628 CJ Delft, Netherlands}
	\author{{\L}ukasz Cywi\'{n}ski}
	\affiliation{Institute of Physics, Polish Academy of Sciences, al.~Lotnik{\'o}w 32/46, PL 02-668 Warsaw, Poland}
	\author{Jaros\l aw K. Korbicz}
	\affiliation{Centre for Theoretical Physics, Polish Academy of Sciences, al.~Lotnik{\'o}w 32/46, PL 02-668 Warsaw, Poland}
	
	\begin{abstract}
	Quantum-to-classical transition still eludes a full understanding. Out of its multiple aspects, one has recently gained an increased attention - the appearance of objective world out of the quantum.  One particularly idea is that objectivity appears thanks to specific quantum state structures formation during the evolution, known as Spectrum Broadcast Structures (SBS). Despite that quite some research was already performed on this strongest and most fundamental form of objectivity, its practical realization in a concrete physical medium has not been analyzed so far. In this work, we study the possibility to simulate objectivization process via  SBS formation using widely studied Nitrogen-Vacancy centers in diamonds. Assuming achievable limits of dynamical polarization technique, we show that for high, but experimentally viable polarizations ($p>0.5$) of nuclear spins and for magnetic fields lower than $\approx \! 20$ Gauss the state of the NV center and its nearest polarized environment approaches reasonably well an SBS state.
	\end{abstract}
	
	\date{\today}
	
	\maketitle

	\section{Introduction}
The central spin model - of a two-level system interacting with many other spins - is not only a paradigmatic model of decoherence \cite{Zurek_PRD82,Cucchietti_PRA05}, but it has been highly relevant for description of dephasing of many kinds of semiconductor-based electron spin qubits interacting with nuclear spins \cite{Hanson_RMP07,Coish_pssb09,Cywinski_APPA11,Urbaszek_RMP13,Chekhovich_NM13}. Dynamics of nuclear-induced decoherence has been understood to a very large degree for many kinds of spin qubits interacting with nuclear environments consisting of between $\sim \!10^2$ to $\sim \! 10^{6}$ nuclei \cite{Yang_RPP17,Cywinski_APPA11,Chekhovich_NM13}. For nitrogen-vacancy (NV) centers in diamond \cite{Dobrovitski_ARCMP13,Rondin_RPP14} we are dealing with rather small environment of few hundreds of spins. The spin qubit based in this center has been extensively studied theoretically and experimentally in order to characterize its spin environment (both natural, consisting of spins of $^{13}$C isotope \cite{Zhao2011,Zhao2012,Kwiatkowski_PRB18}, and artificially modified by putting organic molecules on top of the diamond \cite{Staudacher_Science13,Lovchinsky_Science16}) by analyzing the time-dependence of dephasing of an appropriately driven qubit \cite{Degen_RMP17}. 
Most importantly for us here, a large progress has been made in controlling the state of at least a part of this environment - up to a few tens of nuclear spins most strongly coupled to the central spin (the qubit) - and using the center to sense the state of at least some of these environmental spins. Having a well-tested theoretical model of open system dynamics for NV centers interacting with their nuclear environment \cite{Zhao_PRB12,Yang_RPP17}, one can shift the focus from the process of qubit's loss of coherence, to the possibly accompanying processes of modification of environmental state due to interaction with the qubit.

In the process of decoherence, qubit can leave traces of its presence in the environment. If we treat the environment as a channel through which many observers can acquire information about the qubit, then we can try to find how objectively this information is proliferated. Objectivity, as an important part of the quantum-to-classical transition, has been recently receiving a growing research attention, see e.g. \cite{Le20, Paternostro20, Maniscalco20, Oliveira19, ZurekJelezko, Pan19, Ciampini18}  for recent developments. The problem of objectivity, i.e. how to explain a robust objective world of everyday experience from quantum postulates, was first raised by W.H. Zurek and collaborators \cite{Ollivier04, ZurekNature}, who realized that decoherence alone is not enough, as nothing a priori guarantees that during its course information about the decohering system will make it into its environment in many copies accessible to independent observations - a prerequisite of objectivity. There have been proposed several approaches to the problem, with quantum Darwinism \cite{Ollivier04, ZurekNature} being the first and the most popular one, followed by Spectrum Broadcast Structures (SBS) 
\cite{KHH, HKH, Tuziemski15, Mironowicz17, measurements, GPT, Tuziemski19} and strong quantum Darwinism \cite{Le19}. All the approaches can be viewed as extensions of theory of decoherence, in which  one is interested not only in the system's state but also in what information about it, leaks into the environment (assumed to be a compound quantum system itself). The first and the last approaches study the behavior of quantum mutual information between the system and the parts of the environment, while SBS concerns directly the structure of quantum states. 
	The rigorous relationships among them have been shown in \cite{Le19}: SBS and strong Darwinism both imply the original quantum Darwinism but not vice versa since the original quantum Darwinism is in a sense too weak a condition for classicality as it can still allow for information not accessible locally (via quantum discord) (see also \cite{LePRA, Roads}). The difference between SBS and the strong quantum Darwinism is in turn rather small, with the latter allowing for a bit more general, correlated structure of the environment (a fact already noted in \cite{HKH} and thus both can be regarded as largely equivalent. A more detailed account of different approaches can be found in \cite{Roads}. Since strong quantum Darwinism requires calculations of quantum discord, which are in general difficult, we will use in this work SBS formation as an indicator of objectivity.
	
Let us briefly recall \cite{KHH,HKH} that SBS are the following multipartite state structures:
		\begin{equation}\label{SBS}
			\densmat_{Q:E_{obs}}=\sum\limits_{m}p_m\ket{m^Q}\bra{m^Q}\otimes \densmat_m^{E_1}\otimes...\otimes\densmat_m^{E_{N}}
		\end{equation}
		where $E_{obs}$ is the observed part of the environment, $\ket{m^Q}$ are so called pointer states to which the central system decoheres \cite{Zurek_PRD81,Zurek_PRD82} and the system state conditional density matrices of environmental parts must have mutually orthogonal supports and as a result be perfectly distinguishable:
		\begin{equation}
			\densmat_{m}^{E_k}\perp\densmat_{m'}^{E_k}
			\label{eq:perp}
		\end{equation}
		It is straightforward to see that due to (\ref{eq:perp}) each fragment of the environment perfectly encodes the same pointer state index $m$ and it is locally measurable without any disturbance (on average) to the whole state (\ref{SBS}). But this is nothing else than an operational form of objectivity \cite{ZurekNature} or to be more precise intersubjectivity \cite{Mironowicz17}. Surprisingly, the converse is also true \cite{HKH}: SBS (possibly generalized to correlated environments \cite{Le19}) is the only state structure compatible with the quoted notion of objectivity. 
		Interestingly, in some recent experiments \cite{Ciampini18,Pan19} what in fact has been observed is a formation of the SBS states \cite{Roads}.

We will discuss the formation of SBS structures in the experimentally widely investigated system of nitrogen-vacancy (NV) centers in diamond. 
It is worth mentioning at this point that recently a state of the art experiment has been performed \cite{ZurekJelezko}, reporting an emergence of a (somewhat reduced) form of quantum Darwinism in NV system. While undoubtedly pioneering and of a great importance, in the light of the above discussion it represents rather the first step in using NV systems as 'simulators of objectivity'. In particular SBS represents the strongest form of objectivity and it is an interesting question if NV centers can simulate it.

The electronic energy levels of these centers lie in the bandgap of diamond, and the ground state manifold of the NV center corresponds to spin $S=1$ (e.g. \cite{Dobrovitski_ARCMP13}) system.
The selection rules for coupling of photons to relevant transitions allow for optical initialization of spin-polarized state within the ground state manifold. By choice of microwave resonant drive between two out of three possible spin levels, one can experimentally define a qubit. Additionally, very weak spin-orbit coupling causes that the NV center decoherence is caused mostly by coupling to the environment formed out of $^{13}$C nuclear spins randomly uniformly distributed through the lattice structure \cite{Zhao_PRB12}. Natural concentration of those nuclei is around 1.1 \%, so the environment consists of rather sparsely distributed spins, the spatial arrangement of which does not reflect the periodicity of underlying crystal lattice. These spins are coupled to the NV center qubit, and also among themselves, via anisotropic dipolar interactions, whose power-law ($1/r^3$) decay with distance makes nearby spins much more strongly coupled than the remote ones, but does not allow for treating the interaction as having finite range. This, together with the sparsity of the environment, means that the coupling constants in the Hamiltonian for each NV-environment system are specific to the given spatial arrangement of nuclei (``spatial realization of environment''). The experiments are most often done at finite magnetic fields, so the environmental spins undergo Larmor precession. The resulting  dynamics due to this precession, qubit-nuclear coupling, and inter-nuclear interactions strongly depends on the value of magnetic field. In the case on which we are focusing here - that of freely evolving qubit not subjected to any kind of dynamical decoupling that prolongs its coherence time \cite{Yang_RPP17} - complete dephasing of the qubit occurs on timescale on which inter-nuclear interactions play no role \cite{Zhao_PRB12}. However, the SBS emerge only after a time of decoherence caused by a part of the environment \cite{KHH}, which is longer than the time of decoherence due to the whole environment. Consequently, we will pay here careful attention to relevant  timescales, in order to maintain the vailidity of approximation of treating the nuclear spins as mutually non-interacting.

For typically used values of magnetic field and temperature, the nuclear density matrix is very close to a completely mixed one. SBS cannot form with an initially completely mixed state of environment for a simple reason that such environment is completely ignorant to any information about the system and there is no chance for the condition (\ref{eq:perp}) to be fulfilled (see e.g. \cite{KHH,Roszak_Korbicz_PRA19}). However, there has been a recent progress in generation of so-called dynamical nuclear polarization (DNP) of the nuclear spins most strongly coupled (i.e.~the closest) to the NV center \cite{London2013,Fischer2013a,Fischer2013b,Pagliero2018,Wunderlich2017,Alvarez2015,King2015,Scheuer2017,Hovav2018,Schwartz2018}. Consequently, we focus here on the case in which such a DNP is present, and we analyze the emergence of Spectrum Broadcast Structures as a function of polarization of the nuclei, and the size of the polarized fraction of the environment. A novel aspect of our SBS analysis is the inclusion of a non-trivial dynamics for the spin environment. This is an important generalization of the spin-spin models studied so far \cite{Mironowicz17, Mironowicz18}. In those studies, the environment self-Hamiltonian was completely neglected, leading to a very simplified and a rather academic model. Here we present a more realistic one.

The work is organized in the following way. In Section \ref{sec:Model}, we shall present the Hamiltonian for an NV center interacting with an environment of $^{13}$C nuclei. 
In Section \ref{sec:SBS} we first study a  general model of SBS formation in spin systems with a non-trivial environment dynamics. We also discuss the model of the nuclear environment. We then apply the model to the situation when the nuclear environment interacts only with the central qubit, i.e. there are no direct interactions between the bath spins. 
In Section \ref{sec:res} we perform numerical analysis of the model described in the prior sections, showing the regime of SBS formation under realistic conditions for NV centers in diamond with natural concentration of $^{13}$C nuclei. Concluding remarks are presented in Section \ref{sec:conclude}.

	\section{The Model}\label{sec:Model}
	The system of the NV center and its nuclear environment is described by a pure dephasing Hamiltonian:
	\begin{equation}
		\HH=\HH_Q+\HH_E+\Sz\hat{V}
	\end{equation}
	where $\HH_Q$ is the Hamiltonian of the qubit, $\HH_E$ of the environment, $\Sz$ is the $z$ component of center's spin (with $z$ axis being determined by the vector connecting the nitrogen and the vacancy), and $V$ is the environmental operator that couples to the qubit. A special feature of the NV center is that its low-energy subspace relevant for qubit physics is that of spin $1$, so that $\Sz$ has eigenvalues $m\! =\ -1$, $0$, $1$. The qubit's Hamiltonian is
	\begin{equation}
		\HH_Q=\Delta_0\Sz^2+\bar\Omega\Sz,
	\end{equation}
	where $\Delta_0 = $2.87 GHz is the zero-field splitting, between the $m\! =\! 0$ and $m\! =\! \pm 1$ states, and $\bar\Omega=\gamma_e B$ GHz is the Zeeman splitting between $m_s=\pm 1$ levels due to external magnetic field $B$. Gyromagnetic ratio of the electron is equal to $\gamma_e=28.07$ GHz/T. Note that the $B$ field is assumed to be parallel to the NV center quantization axis.
	There is a freedom of choosing any $2$ out of $3$ energy levels to define the qubit. Here we focus on the most popular (due to experimental ease of manipulation) choice of qubit based on $m\! = \! 0$ and $m\! = \! 1$ levels.

The environmental Hamiltonian consists of the Zeeman splittings term and the inter-nuclear interactions:
	\begin{equation}\label{HE}
		\HH_E=\sum_i \omega_i \I^{i}_z+\HH_{int},
	\end{equation}
	where $\omega_i=\gamma_{^{13}C} B$ MHz is the Zeeman splitting with gyromagnetic ratio of $^{13}$C nuclei, $\gamma_{^{13}C}=10.71$ MHz/T,  and $\I^{(i)}_z$ is the $z-$axis spin operator of the $i$-th nuclear spin. 
	There are two mechanisms of electronic spin-nuclear spin coupling: Fermi contact interaction, which is proportional to the overlap of the electronic wavefunction at the position of a nucleus $A_{Fermi}\propto|\psi_e(\mathbf{r}_{i})|^2$, and dipolar interaction. The former is negligible for nuclei farther away than $0.5$ nm from the center \cite{Gali2008} as the wavefunction is highly localized for deep defects. Within this radius, for $\approx 50$\% possible realizations of the environment there will be no spinful nuclei. Keeping in mind the post-selection of spatial realizations of the environment that needs to be done, we will focus from now on only on dipolar qubit-nuclear spin couplings.

For magnetic fields of interest here, the order of magnitude of qubit energy splitting is determined by the zero-field splitting $\Delta_0$, which is much larger than the nuclear energy scales (Zeeman splittings, dipolar interactions). Consequently, the qubit and its environment cannot exchange energy, i.e.~we are dealing with pure dephasing, and we can neglect terms $\sim \hat{S}_x,\hat{S}_y$ in the qubit-nuclear coupling, which is therefore given by
	\begin{equation}
		\Sz \hat{V}=\sum\limits_k\sum\limits_{j=x,y,z} \Sz A^{j}_{k}\hat{I}^k_j \,\, ,
	\end{equation}
	where $j\! =\! x$, $y$, $z$ enumerates directions of spin operators, $k$ - nuclear spins interacting with the qubit and $A^{j}_k$ are given by:
	\begin{equation}
		A^{j}_k=\frac{\mu_0 \gamma_e\gamma_{^{13}C}}{4\pi}\frac{\hat{\mathbf{z}}\cdot\hat{\mathbf{j}}}{|\mathbf{r}_k|^3}-\frac{3(\hat{\mathbf{z}}\cdot \mathbf{r}_k)(\hat{\mathbf{j}}\cdot\mathbf{r}_k)}{|\mathbf{r}_k|^5},
		\label{eq:Ajk}
	\end{equation}
	where $\mu_0$ is the magnetic permeability of vacuum, $\mathbf{r}_k$ is a displacement vector between nitrogen and nucleus $k$ and the gyromagnetic ratios $\gamma_e$ and $\gamma_{^{13}C}$ are defined above. 
	
For qubit based on $m\! =\! 0$ and $1$ levels that we consider here, the qubit-environment coupling is then given by $\ket{0}\bra{0} \hat{V}_0+\ket{1}\bra{1}\hat{V}_1$ in which  
\begin{equation}\label{Vm}
\hat{V}_{m} = m\sum\limits_k\sum\limits_{j=x,y,z}  A^{j}_{k}\hat{I}^k_j,
\end{equation}
so that $\hat{V}_0 \! =\! 0$.
 Consequently, the evolution operator of the whole system can be written as:
	\begin{equation}\label{Uint}
		\hat{U}(t)=\ket{0}\bra{0}\otimes \U_0(t)+\ket{1}\bra{1}\otimes \U_1(t) \,\, ,
	\end{equation}
	where the conditional evolution operators are given by
	\begin{equation}\label{Um}
		\U_m(t)=\exp[-it(\HH_E+\hat{V}_m)] \,\,.
	\end{equation}
We are working in the qubit rotating frame, so that the energy splitting of the qubit $\Omega+\Delta_0$ is removed.
	
\begin{figure}[ht]
		\begin{center}
			\includegraphics[trim={0cm 0cm 0cm 8cm},clip,width=0.5\textwidth]{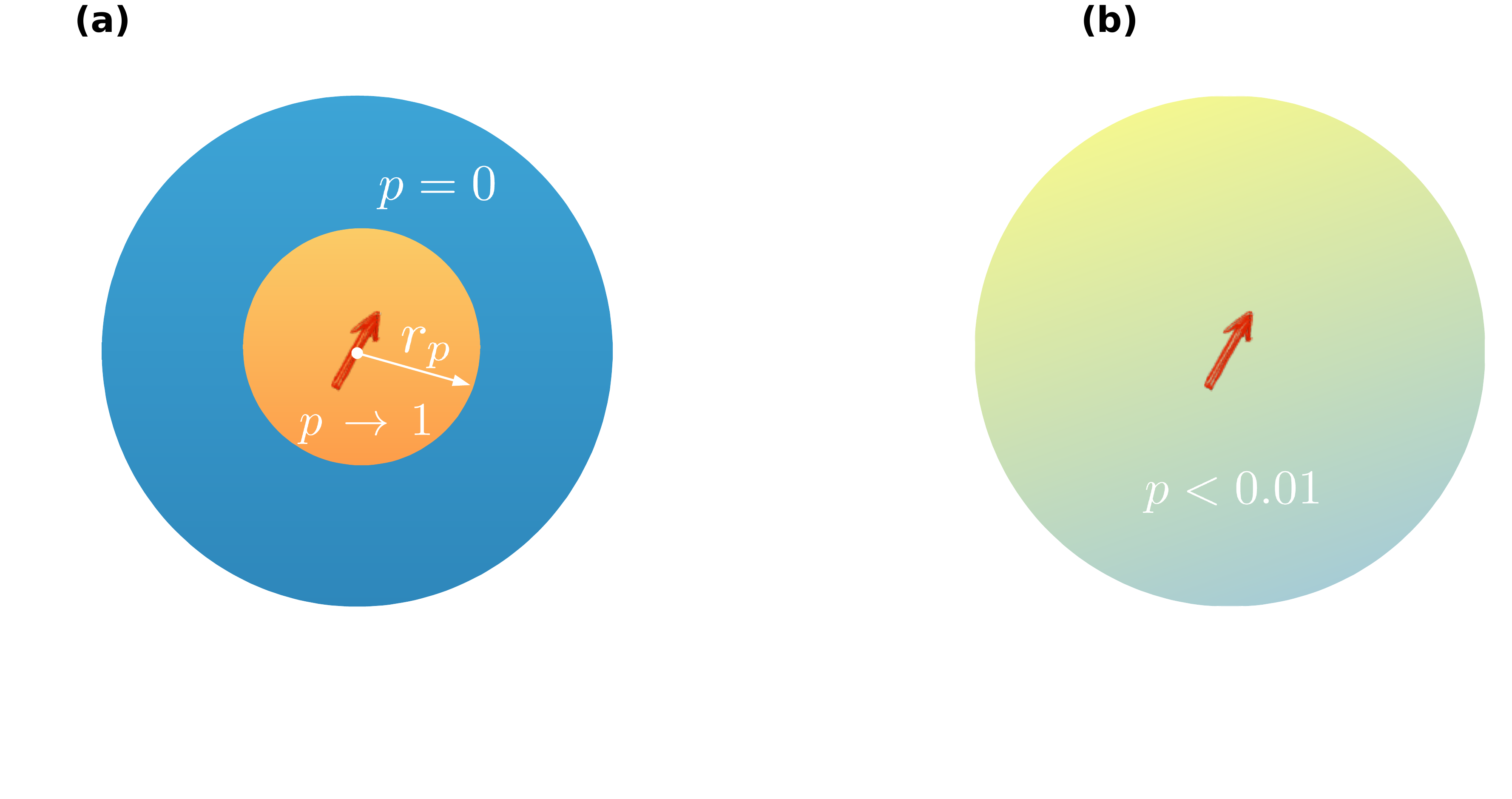}
		\end{center}
	
		\caption{Experimental strategies for nuclear polarization in NV centers. Panel (a) depicts a situation when strongly coupled nuclei, located up to a radius $r_p$ from the NV center, are polarized using dynamic nuclear polarization (DNP). Panel (b) shows the case for achieving a polarization in high magnetic fields and cryogenic temperatures to increase the ratio between Zeeman splitting and temperature, which corresponds to thermal polarization $p_i=\tanh(\omega_i/2 k_{B}T)$. In this work we study case (a), unless otherwise stated.}
		\label{fig:pol}
	\end{figure}

Having discussed the dynamics, let us now discuss the initial conditions. If the spin environment is non-polarized, the interaction (\ref{Uint}) will not lead to any information recording in the environment, only to dephasing \cite{Roszak2015}. Therefore we will consider here partially polarized environments. Specifically, we focus on environments in which nuclear spins within some distance from the qubit are polarized, as such an environmental state can be prepared by repetition of appropriate manipulation protocols \cite{London2013,Fischer2013a,Fischer2013b,Pagliero2018,Wunderlich2017,Alvarez2015,King2015,Scheuer2017,Hovav2018,Schwartz2018} on the qubit and the nuclei, and the efficiency of polarization scales with the magnitude of qubit-nuclear coupling. Such an initial state of the environment is illustrated in Fig.~\ref{fig:pol}. Another strategy for nuclear polarization is to put the diamond crystal in cryogenic temperatures and apply high magnetic field, resulting in uniform polarization after the nuclear spins reach thermal equilibrium with the lattice, see Fig.~\ref{fig:pol}b. However, equilibrium polarization for temperature of a few tens of milikelvins and at $B\sim 1$T is $p\sim 10^{-2}$. Below we will see that such a polarization is not enough to support formation of the SBS. 
	
	\section{Dynamics of the SBS formation}\label{sec:SBS}
	\subsection{General considerations}
As explained in the Introduction, our method  is based on direct studies of the quantum state as the most fundamental carrier of information. In particular, we are interested if there are regimes such that a joint state of the central qubit and some of its nuclear environment approaches the SBS structure (\ref{SBS},\ref{eq:perp}), signalizing that the state of the qubit acquired a certain operational objective character during the evolution as explained in detail in \cite{KHH,HKH}. As in the previous SBS studies, e.g. in \cite{KHH, Tuziemski15, Mironowicz17, Mironowicz18}, our method is the following. First, since we are interested not only in the state of the qubit alone but in how it is correlated with some of its environment, we cannot trace all of the environment as it is normally done. Instead, we divide the environment $E$  into two parts: The one we are interested in (say observed), denoted symbolically $fE$ and containing $fN$ spins, $0\le f \le 1$, and the one that pass unobserved and can be traced out, denoted $(1-f)E$ and containing the rest of the $(1-f)N$ nuclei. In terms of experimental capabilities, one may think of DNP as a form of environment separation. As described above, high degree of polarization can be reached for only a few nuclear spins closest to the NV center. Control and observation of polarized fragment of the environment can be realized by measurement $\langle \hat{\sigma}_y \rangle$ of the qubit as a function of total evolution time, which is zero when the environment is completely mixed during evolution of a qubit initialized in  eigenstate of $\hat{\sigma}_x$.

The main object of our study is what we call a partially reduced state:
		\begin{equation}
			\densmat_{Q:fE}(t)=\Tr_{(1-f)E}\densmat_{Q:E}(t),
		\end{equation}
		obtained by tracing out only the unobserved part of the environment, $(1-f)E$, from the global qubit-environment state $\densmat_{Q:E}(t)$ evolving under (\ref{Uint}).  The check for SBS structure then proceeds in two steps \cite{Mironowicz17}: (i) first check if dephasing takes place and the partially reduced state approaches the form Eq. \eqref{SBS}; 
		ii) check if the conditional environment states satisfy (\ref{eq:perp}). The first condition, dephasing, is fairly standard and we will use well-known results, scaled down however to a part of the environment rather than the whole. When it comes to the second condition, out of the several available measures of state distinguishability (\ref{eq:perp}) [cite Fuchs], we use the state fidelity: 
\begin{equation}\label{F}
\F(\hat\rho,\hat\sigma)=\Tr(\sqrt{\hat\rho}\hat \sigma\sqrt{\hat \rho}) 
\end{equation}
for the ease of work. In any case, we are interested only in $\F=0$, which is equivalent to (\ref{eq:perp}). It can happen that a state of a single environment nucleus is changed too little during the evolution (\ref{Uint}) to approach (\ref{eq:perp}), but when we consider groups of nuclei, their joint states can come close to satisfying (\ref{eq:perp}). This can be viewed as a kind of "information concentration".  Anticipating such situation, we introduce following \cite{KHH} a further coarse-graining of the observed environment $fE$ into $M$ groups, called macrofractions, each of a size $\mu N = (f/M) N$  (equal sizes are for our convenience only). The approach to SBS is then mathematically equivalent to simultaneous vanishing of the decoherence factor due to $(1-f)E$ and all the pairwise fidelities calculated between the states of the macrofractions \cite{Mironowicz17}. 
We note that for pure states the fidelity (\ref{F}) becomes just the overlap $F(\psi,\phi)=|\langle \psi|\phi\rangle|^2$.

\begin{figure}[ht]
	\begin{center}
		\includegraphics[trim={0cm 3cm 18cm 3cm},clip,width=0.3\textwidth]{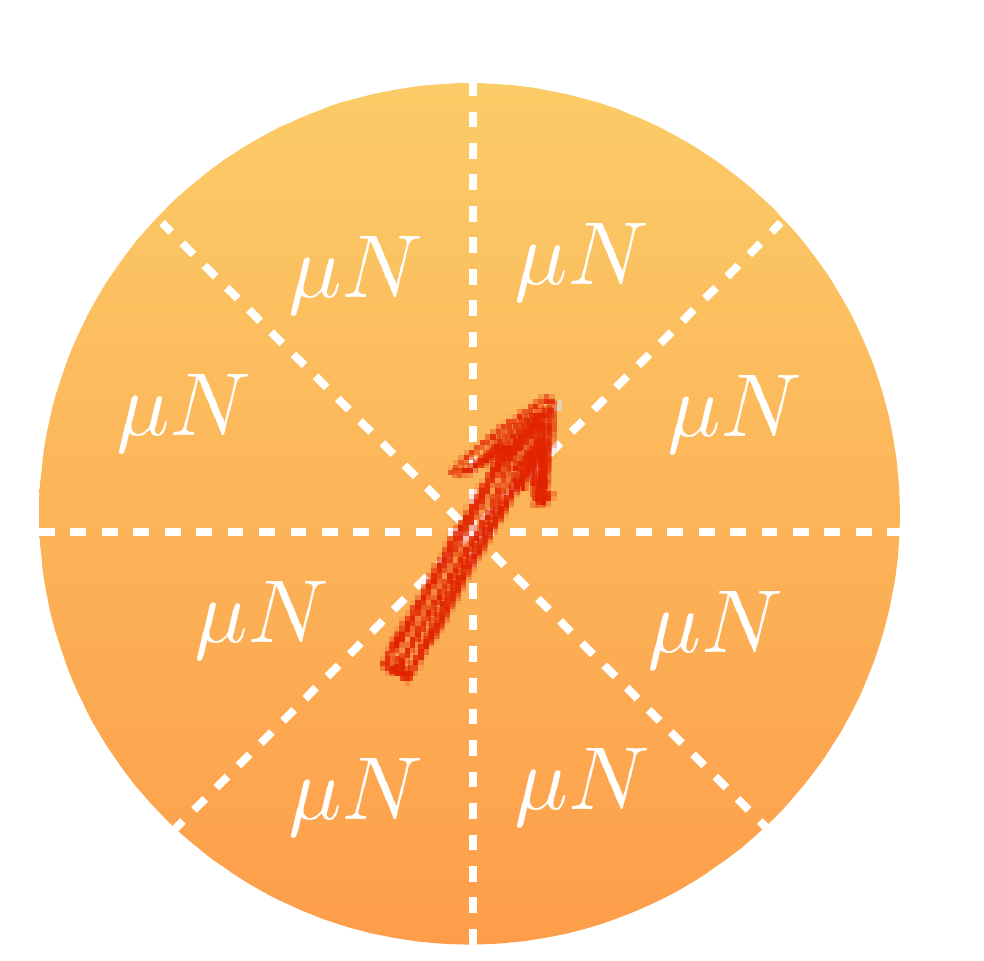}
	\end{center}
	
	\caption{Schematic representation of the coarse-graining of the observed part of the environment, $fE$, into macrofractions containing $\mu N$ spins each. 
This helps achieving "information concentration", defined here  by the perfect distinguishability of the states (\ref{eq:perp}), and thus helps approaching  SBS states. This is more general situation than just considering each environmental spin individually.}
	\label{fig:sbs_scheme}
\end{figure}

The concrete setup studied here will be the DNP setup of Fig.~\ref{fig:pol}a with the following identifications:
\begin{itemize}
\item The central system is the NV qubit, defined by the $\ket{m^Q=0}$ and $\ket{m^Q=1}$ states, which constitute the pointer basis. We are seeking if during the interaction with the environment, the decohered state of the qubit becomes objective via a creation of the SBS state (\ref{SBS}). 
\item The observed part of the environment, $fE$, will be the DNP spins within the radius $r_p$ from the NV center
\item The observed part will be further divided into several, equal size, macrofractions (see Fig.~\ref{fig:sbs_scheme}). They represent parts of the environment accessible for independent observers.
\item The weakly polarized part of the environment past the radius $r_p$ carries vanishingly small amount of information about the qubit and thus this will be the unobserved part, $(1-f)E$, subsequently traced over.
\end{itemize}

We assume that the initial state of the qubit and all the nuclei is initially completely uncorrelated:
\begin{equation}
		\densmat_{Q:E}(0)=\densmat_{Q}(0)\otimes\densmat_E(0)=\densmat_Q(0)\bigotimes\limits_k\densmat_E^k(0) \,\, ,
	\end{equation}
	where $k$ enumerates nuclei in the bath, and the state of a single nucleus is given by 
	\begin{equation}\label{rho_init}
		\densmat_E^{k}(0)=\frac{1}{2}(\mathbbm{1}+p_k\hat{\sigma}_z^{(k)}) \,\, ,
	\end{equation}
	in which $p_k$ is the initial polarization degree. When $p_k \! = \! \pm 1$, the state is pure, and when $p_k \! = \! 0$, the state is fully mixed. In other words, only spins affected by DNP, thus, forming the observed environment, will correspond to $p_k\!\neq\! 0$ and for the unobserved part, we assume $p_k\!=\!0$, corresponding to room temperature - typical conditions for experiments with NV centers.

	Anticipating the irrelevance of inter-nuclear interactions, 
the total Hamiltonian reads:
	\begin{equation}\label{ham}
	\hat{H} = (\Delta_0+\bar\Omega)\ket{1}\bra{1} + \sum_{k}\omega_k \hat{I}^{k}_z + \ket{1}\bra{1}\otimes\sum_k \sum_{j=x,y,z} A^{j}_{k}\hat{I}^{k}_j \,\,\ .
	\end{equation}
	This  Hamiltonian allows for a correct description of decoherence of a freely evolving NV center spin qubit \cite{Zhao_PRB12}, which has also been used for interpretation of experimental signal from such an NV center \cite{Liu2012}.
From the point of view of objectivity and SBS studies, the above Hamiltonian is an important generalization of the previously studied spin-spin models \cite{Mironowicz17,Mironowicz18}.

	When the central qubit is initialized in a pure  superposition of pointer states, i.e.~in
\begin{equation}
\densmat_{Q}(0)=\ket{\phi}_{Q} \bra{\phi},\ \ket{\phi}_{Q}\! =\! c_0\ket{0}+c_1\ket{1}, 
\end{equation}
the evolution of total system, governed by (\ref{Uint}), is given by:  

	\begin{equation}
			\densmat_{Q:E}(t)=\sum\limits_{m,m'=0,1} c_mc_{m'}^*\ket{m^Q}\bra{m'^Q}\bigotimes_{k=1}^{N}\U^k_m\densmat_E^k(0)\U_{m'}^{k\dagger} 
	\end{equation}

\subsection{Analytical results - decoherence factor}

Once we trace out the unobserved part of the environment $(1-f)E$, the partially reduced density matrix becomes:
\begin{align}
			&\densmat_{Q:fE}(t)=\sum\limits_{m=0,1}|c_m|^2\ket{m}_Q\bra{m}\bigotimes_{k=1}^{fN}\U^k_m\densmat^k_{E}(0)\U^{k\dagger}_m+\nonumber\\
			&+\left(c_mc_{m'}^*\gamma_{mm'}(t)\ket{m}_Q\bra{m'}\otimes \bigotimes_{k=1}^{fN}\U^k_m\densmat^k_{E}(0)\U_{m'}^{k\dagger}+c.c.\right)\label{rfE}
\end{align}
	where $\gamma_{mm'}(t)$ is the decoherence factor coming from the unobserved fraction of the environment $(1-f)E$. For the chosen realization of a qubit between $m=0$ and $m=1$ states, this term can be expressed as:
	\begin{equation}\label{full_gamma}
		\gamma(t)=\prod\limits_{k=1}^{(1-f)N}\gamma_{k}(t)= \prod\limits_{k=1}^{(1-f)N}\Tr\left(\U^{k}_0\densmat_{k}^{E}(0)\U^{k\dagger}_1\right),
	\end{equation}
	where the single nucleus decoherence factor $\gamma_{k}(t)$ reads by (\ref{Vm},\ref{Uint},\ref{Um}):
	\begin{align}
		&\gamma_k(t)=\cos\frac{\omega_k t}{2}\cos\frac{\Omega_k t}{2}+\frac{A^z_k+\omega_k}{\Omega_k}\sin\frac{\omega_k t}{2}\sin\frac{\Omega_k t}{2} \nonumber\\
		& +i p_k \left(\cos\frac{\Omega_k t}{2}\sin\frac{\omega_k t}{2}-\frac{A^z_k+\omega_k}{\Omega_k}\cos\frac{\omega_k t}{2}\sin\frac{\Omega_k t}{2}\right)\,\, , \label{eq:gamma}
	\end{align}
	in which $\Omega_k=\sqrt{\left(A^{\perp}_{k}\right)^2+(\omega_k+A^{z}_{k})^2}$ and\\ $A^{\perp}_{k}=\sqrt{\left(A^{x}_k\right)^2+\left(A^{y}_k\right)^2}$. The modulus is given by:
\begin{eqnarray}
&&|\gamma_k(t)|^2=\left[1-(1-p_k^2)\sin^2\frac{\omega_k t}{2}\right]\cos^2\frac{\Omega_k t}{2}\nonumber\\
&&+\left(\frac{A^z_k+\omega_k}{\Omega_k}\right)^2\left[1-(1-p_k^2)\cos^2\frac{\omega_k t}{2}\right]\sin^2\frac{\Omega_k t}{2}\nonumber\\
&&+\frac{A^z_k+\omega_k}{2\Omega_k}(1-p_k^2)\sin(\omega_k t)\sin(\Omega_k t).
\end{eqnarray} 
A general expression for decoherence factor when qubit is defined between $m$ and $m'$ states can be found in Appendix A.

We now have to estimate the product (\ref{full_gamma}) with the factors given by (\ref{eq:gamma}). Analytical studies are possible only under some simplifications. 
The most universal one is the short-time limit $\Omega_k t\ll 1$, which also implies $\omega_k t \! \ll \! 1$, so it can hold only below a certain magnetic field for given timescale of interest. The total decoherence factor then reads:
		\begin{equation}
		\gamma^{weak}(t)\approx\exp\left[ -\left(t/T_{2}^*\right)^2 - i\phi(t)   \right]
		\end{equation}
		 where the dephasing time $T_{2}^{*}$ is defined by  
		\begin{equation}
		(T_2^*)^2= \frac{8}{(1-f)N \langle (A^z)^2 + (A^\perp)^2\rangle} \,\, , \label{eq:T2star}
		\end{equation}
		and the phase shift is given by $\phi(t) = (1-f)N \langle pA^z\rangle t/2$. The averages are defined by:
		\begin{equation}\label{av}
		\langle f(A) \rangle = \frac{1}{(1-f)N}\sum\limits_k^{(1-f)N} f(A_k)  \,\, .
		\end{equation}
As expected, at short times the decoherence factor shows a Gaussian decay but this does not mean that it decays also for larger times. In fact in general it does not for small traced fractions. The further analysis of $\gamma(t)$ will be carried out numerically.

\subsection{Analytical results - conditional states fidelity}\label{sec:fidelity}
	
After the decoherence due to the unobserved part of the environment has taken place, the resulting partially traced state (\ref{rfE}) comes close to the SBS form (\ref{SBS}).
We have to however still check the orthogonality (\ref{eq:perp})  for the conditional states $\hat\rho_m^{E_k}\equiv\hat\rho_m^{k}$, where $\densmat^{k}_m(t)=\U_m(t)\densmat^k_E(0)\U_m^{\dagger}(t)$, cf. (\ref{rfE}).  
		 We will use the state fidelity function (\ref{F}).
		We calculate it, using the fact that all the matrices are $2\times 2$:
		\begin{align}
			&\F_{mm'}=\F\left(\densmat^{k}_m(t),\densmat^{k}_{m'}(t)\right)=\\
			&=\Tr[\densmat^{k}_m(t)\densmat^{k}_{m'}(t)]+2\sqrt{\det[\densmat^{k}_m(t)]\det[\densmat^{k}_{m'}(t)]}.\nonumber
		\end{align}
		For the  qubit based on $m \in\{0,\,+1\}$ levels, considered here, the resulting fidelity for conditional states of the nucleus becomes:
		\begin{equation}
			\F\left(\densmat^{k}_0(t),\densmat^{k}_1(t)\right)=1-\frac{\left(A_k^\perp\right)^2}{\Omega_k^2}p_k^2\sin^2\left(\frac{\Omega_k t}{2}\right),\label{eq:F01}
		\end{equation}
 While single-spin contribution to decoherence, Eq.~(\ref{eq:gamma}), is finite even  when $A^\perp_k \! =\! 0$ (only nonzero $A^{z}_k$ is needed),  for the fidelity between the two conditional states of a single environmental spin to be less than unity, $A^\perp_k \! \neq \! 0$ is necessary. This is a consequence of a simple observation that the environment has to undergo an evolution non-trivially conditioned on the state of the qubit for this fidelity to deviate from unity. We recall that fidelity equals to one iff the states are identical, which is a trivial situation. For the similar reason, the non-polarized limit of $p_k\rightarrow$0 is not interesting either. 

In the studied model of qubit-environment coupling leading to qubit's pure dephasing, and initially pure state of the qubit, the necessary condition for the conditional states to be (approximately) orthogonal at long enough times is appearance of nonzero qubit-environment entanglement at earlier times, in the initial stages of the evolution \cite{Roszak_Korbicz_PRA19}. The condition for the latter is $\densmat_0^k(t)\neq\densmat_1^k(t)$, as shown in \cite{Roszak_PRA15,Roszak_PRA18}.
This motivates why as the observed part of the spin environment we consider only the polarized part. 
These are the nuclear spins inside a ball of radius $r_p$, schematically shown in Figure \ref{fig:pol}, according to experimental state of the art concerning DNP.
		
		As we explained at the beginning of this Section, to increase the chances of satisfying distinguishability condition (\ref{eq:perp}), we perform a coarse-graining of the observed environment $fE$, dividing it into $M$ macrofractions of size $\mu N$ each. Symbolically $fE=\mu E \cup \dots \cup \mu E$. The state of each macrofraction for neglected mutual interactions is just a product $\densmat^{\mu E}_m(t)\equiv\bigotimes\limits_{k\in \mu N}\densmat^{k}_m(t)$ so that using the factorization property of the fidelity we obtain:
		\begin{equation}
			\F_{mm'}^{\mu E}(t)\equiv\F\left(\densmat^{\mu E}_m(t),\densmat^{\mu E}_{m'}(t)\right)=
			\prod\limits_{k\in \mu E} \F\left(\densmat^{k}_m(t),\densmat^{k}_{m'}(t)\right)
			\label{eq:generalFidelity}
		\end{equation}	
	Thus, the fidelity between two qubit-state conditional density matrices of macrofractions is a product of contributions from Eq. \eqref{eq:F01}:
	\begin{equation}\label{Fmac}
		\F^{\mu E}(t)=\prod\limits_{k \in \mu E}\left[1-p_k^2\frac{\left(A_k^\perp\right)^2}{\Omega_k^2}\sin^2\left(\frac{\Omega_k t}{2}\right)\right].
	\end{equation}
	A general expression for a qubit defined between $m$ and $m'$ is much more complicated and can be found in Appendix B.
	
We are now interested when $\F^{\mu E}(t) \to 0$, meaning the condition (\ref{eq:perp}) is satisfied for macrofraction states. The easiest regime for analytical study corresponds to a situation, when: 
\begin{equation}\label{F:approx}
\frac{\left(A_k^\perp\right)^2}{\Omega_k^2}p_k^2\sin^2\left(\frac{\Omega_k t}{2}\right)\ll1 
\end{equation}
for every $k$. This happens when e.g. i) all the members of the macrofraction are weakly coupled to the central spin: 
	\begin{equation}\label{ii}
		\frac{\left(A_k^\perp\right)^2}{\Omega_k^2}\ll1 \Leftrightarrow  \frac{A_k^z+\omega_k}{A_k^\perp}\gg 1 \,\, ,
	\end{equation}
	or when ii) polarization of the observed environment is low, meaning: 
	\begin{equation}
		p_k^2\ll \frac{\Omega_k^2}{\left(A_k^\perp\right)^2}=1+\frac{(A^{z}_k+\omega_k)^2}{\left(A_k^\perp\right)^2} \,\, ,
	\end{equation}
	or when iii) we  consider very short times $\Omega_k t\ll 1$. Then, Eq.~(\ref{Fmac}) can be rewritten as an exponential of a sum of contributions from all the nuclei in the macrofraction:
	\begin{equation}
		\F^{\mu E}(t) \approx\exp\left[-\sum\limits_{k=1}^{\mu N}\frac{\left(A^\perp_k\right)^2}{{\Omega_k^2}}p_k^2\sin^2\left(\frac{\Omega_k t}{2}\right)\right]  \,\, . \label{eq:Fsin}
	\end{equation}
	For short times $\Omega_k t \ll 1$, we can derive an effective timescale of the initial decay of the fidelity:
	\begin{align}
		\F^{\mu E}(t) \approx\exp\left[-\frac{1}{4}\sum\limits_{k=1}^{\mu N}p_k^2 \left(A^\perp_k\right)^2 t^2\right]
		=e^{-\left(\frac{t}{\tau_{\mu}}\right)^2} \,\, , \label{eq:Fexp}
	\end{align}
	where 
\begin{equation}
\tau_{\mu}^{-2}=\frac{1}{4}\mu N\langle p^2 A_{\perp}^2\rangle_{\mu N}, 
\end{equation}
with $\langle\  \cdot\  \rangle_{\mu N}$ denoting the averaging over the macrofraction, similar to (\ref{av}). 
In general, to prove the orthogonalization (\ref{eq:perp}), this short time analysis is of course not enough. As for the behavior of fidelity at long times, we can state the following. Let us assume that:
\begin{equation}
\Omega_k \approx \omega + \frac{\left(A^\perp_k\right)^2}{2\omega}
\end{equation}
(with Zeeman splittings $\omega_k$ assumed to be the same $\omega$ for all the nuclei, implying spatially uniform magnetic field), which holds for $\omega \! \gg \! A^\perp_k$, $A^z_k$. This automatically implies (\ref{F:approx}) via (\ref{ii}) so that we can use (\ref{eq:Fsin}). With $\sigma$ being the standard deviation of distribution of $A^\perp_k$ in the given macrofraction, for $\sigma^2 t/2\omega \! \gg \! 1$ the values of $\sin^2 \Omega_k t/2$ in Eq.~(\ref{eq:Fsin}) are randomly distributed in $[0,1]$. With many spins in the macrofraction, we can replace then $\sin^2 \Omega_k t/2$ terms by their average value of $1/2$, and the fidelity is
		\begin{equation}
		\F_{mm'}^{\mu E}(t \gg \frac{2\omega}{\sigma^2}) \approx \exp\left[-\frac{1}{2\omega^2} \sum\limits_{k=1}^{\mu N}\left(A^\perp_k\right)^2p_k^2  \right] \,\, .
		\end{equation}
If in the macrofraction of interest 
\begin{equation}
\mu N  \frac{\langle p^2 A_{\perp}^2\rangle_{\mu N}}{\omega^2} \! \gg \! 1,
\end{equation} 
which should be treated as a condition for minimal polarization or the number of spins in the macrofraction, then the fidelity decays towards a very small value on timescale that is $\sim \! \omega/\sigma^2$. 

For a qubit with a macrofraction $\mu N$ to form a spectrum broadcast structure, we not only need to meet a condition for mutual orthogonalization for conditional states of the macrofraction, but also decoherence due to the remaining part of the bath. For short times
the ratio of the decoherence and orthogonalization time becomes:
\begin{equation}
	\left(\frac{T^*_2}{\tau_{\mu}}\right)^2 = \frac{\mu}{1-f}\cdot \frac{2\langle p^2(A^{\perp})^2\rangle_{\mu N}}{\langle(A^{z})^2+(A^\perp)^2\rangle_{(1-f)N}}
\end{equation}
Experimental endeavor to measure and control clusters of polarized nuclear spins with NV centers is mostly limited by the decoherence of the NV center. Therefore, if decoherence happened on a longer timescale than orthogonalization, it should be possible to predict formation of SBS, e.g. by state tomography.

\section{Numerical results}\label{sec:res}
\begin{figure}
	\begin{center}
		\includegraphics[width=\columnwidth]{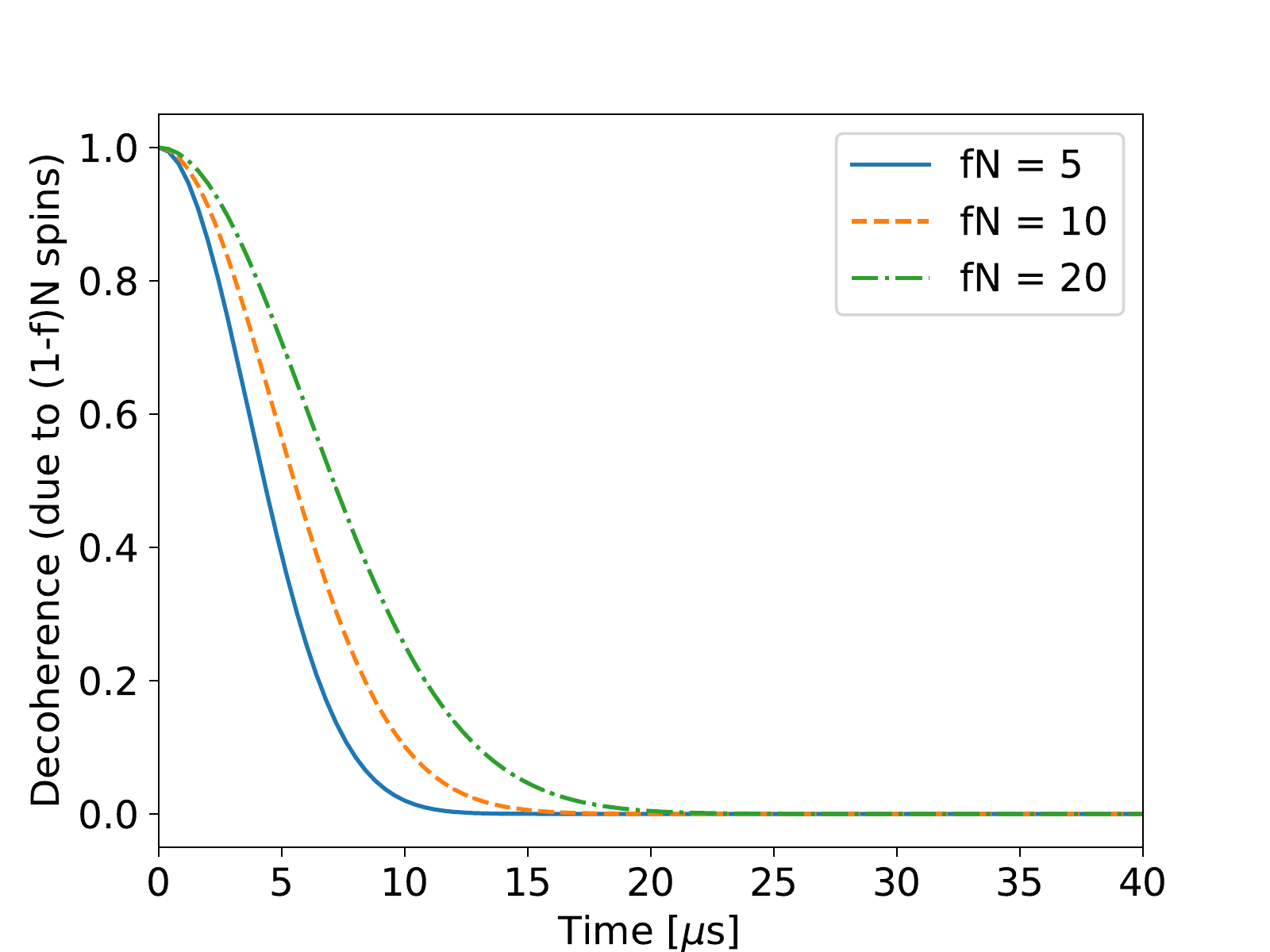}
	\end{center}
\caption{Modulus squared of the decoherence factor (\ref{full_gamma}) as a function of time. The plot corresponds to a single spatial realization of nuclear environment,  labelled as (II).  Due to the fact that unobserved part of environment, used to generate the plot, does not contain by assumption strongly coupled nuclear spins, this Figure can be treated as reference for decoherence timescales independently of realization. The total number of spins in the simulation is $N$=400. Each curve is enumerated by the size of the observed macrofraction  $fE$, rather than that of the unobserved $(1-f)E$, hence the larger the $fN$ the slower the decoherence as less nuclei out of the total $N=400$ contribute to (\ref{full_gamma}). The corresponding state fidelities are shown in column (II) of Fig. \ref{fig:reals}.}
	\label{fig:gamma}
\end{figure}

\begin{figure*}[htbp]
	\begin{center}
		\begin{tabular}{cccc}
			&(I)&(II)&Average over 100 realizations\\
			\textbf{(a)}&&\\
			&	\includegraphics[width=5.5cm]{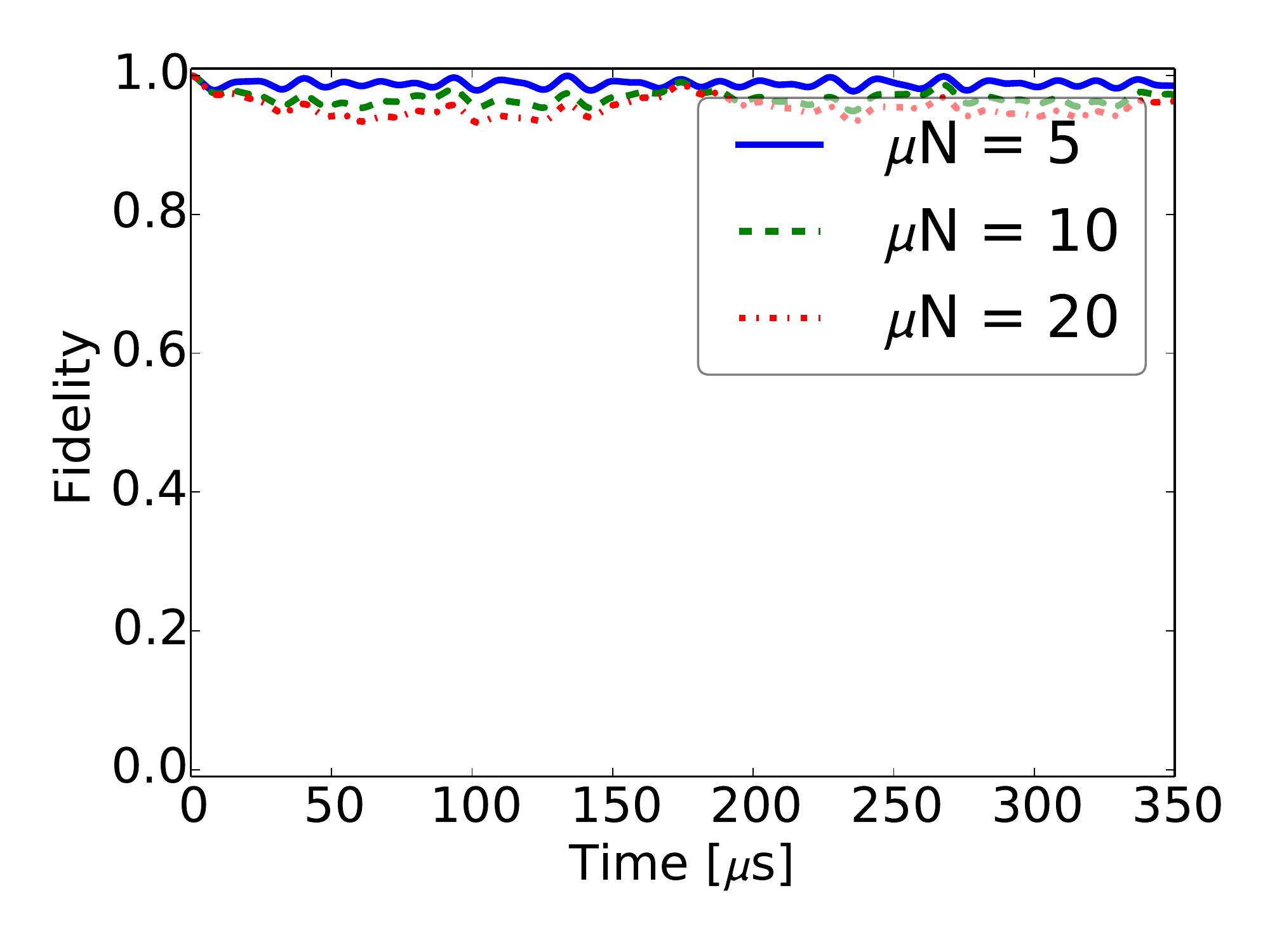}&
			\includegraphics[width=5.5cm]{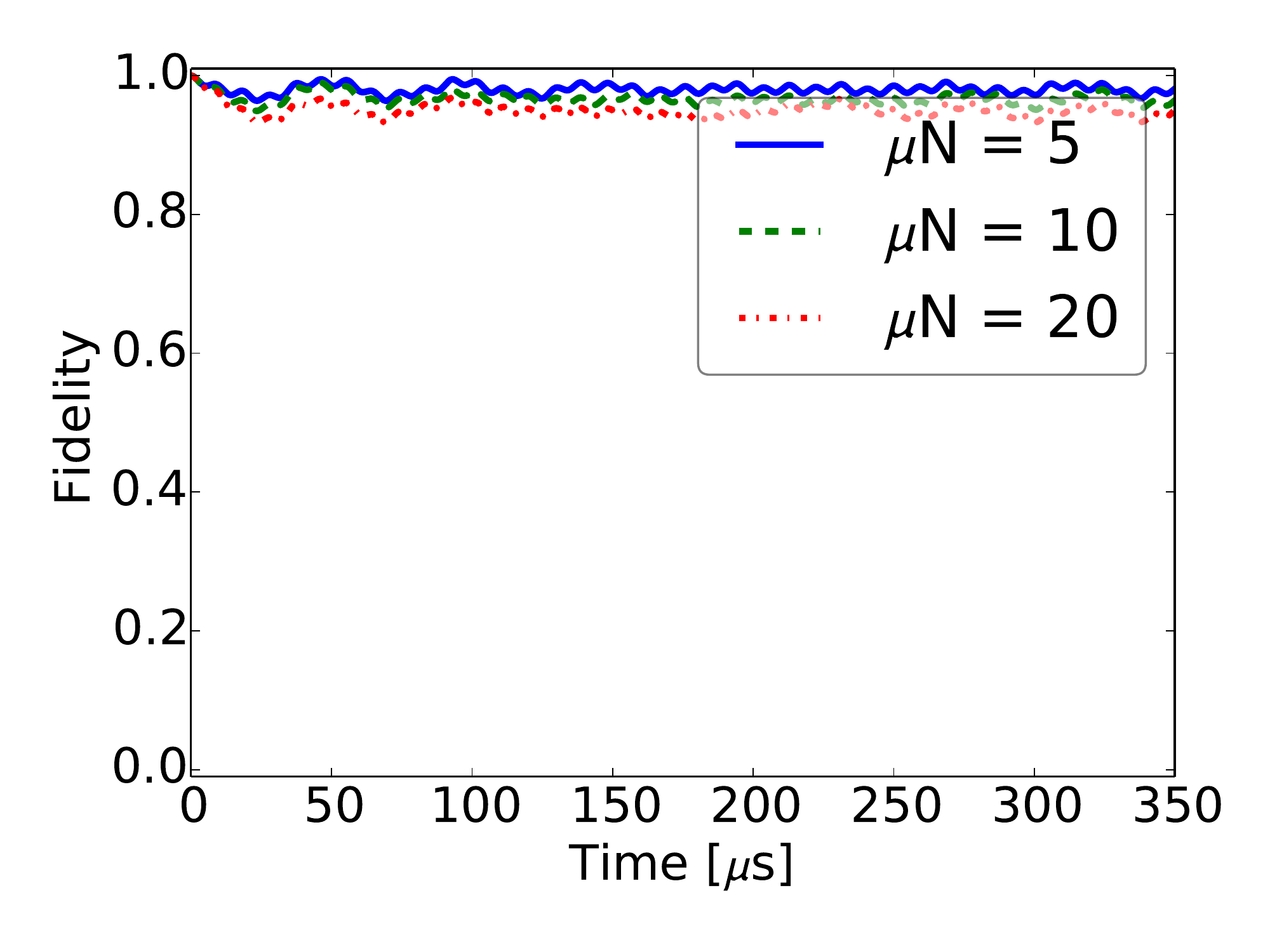}&
			\includegraphics[width=5.5cm]{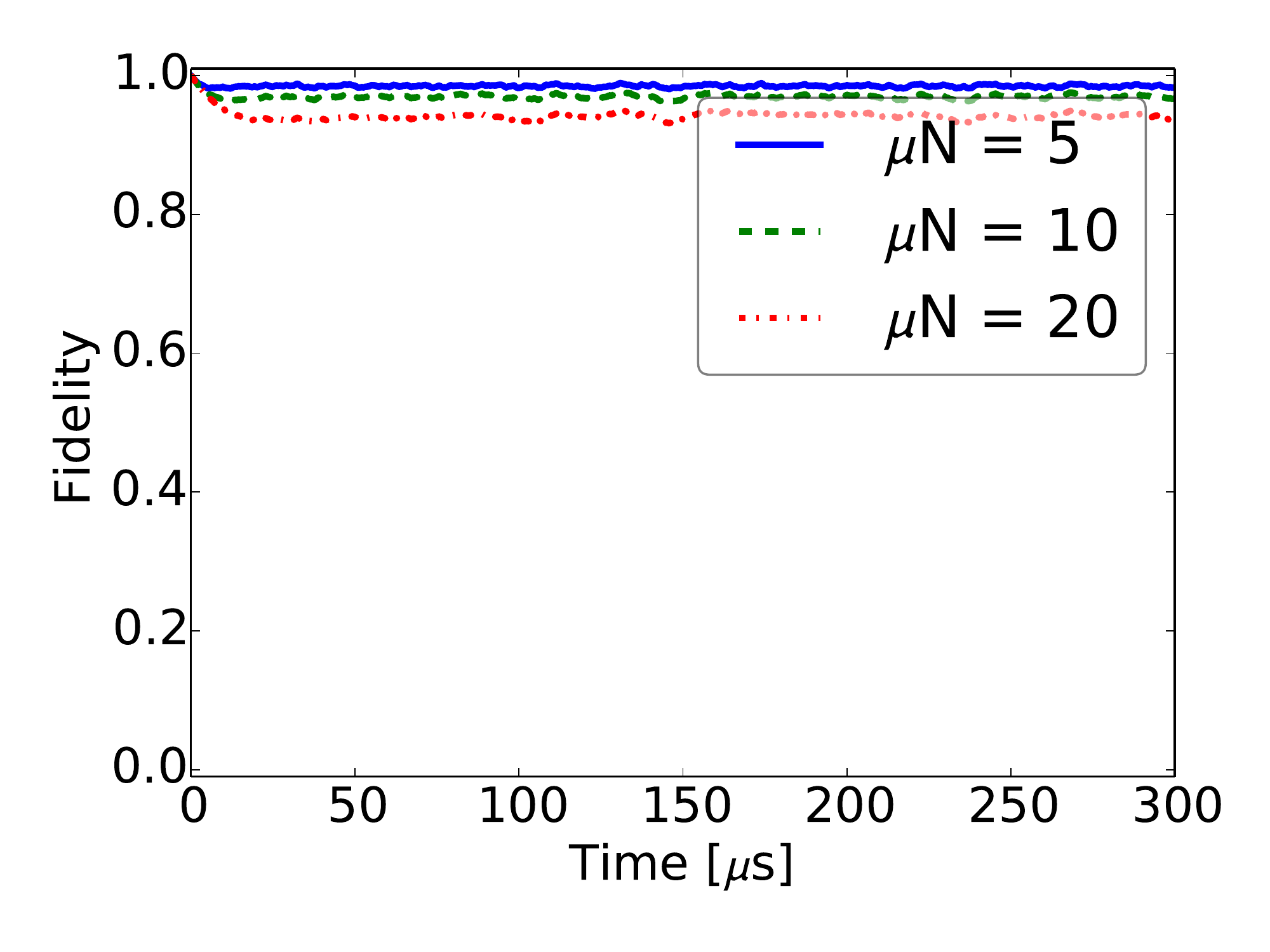}\\
			
			\textbf{(b)}&&\\
			&	\includegraphics[width=5.5cm]{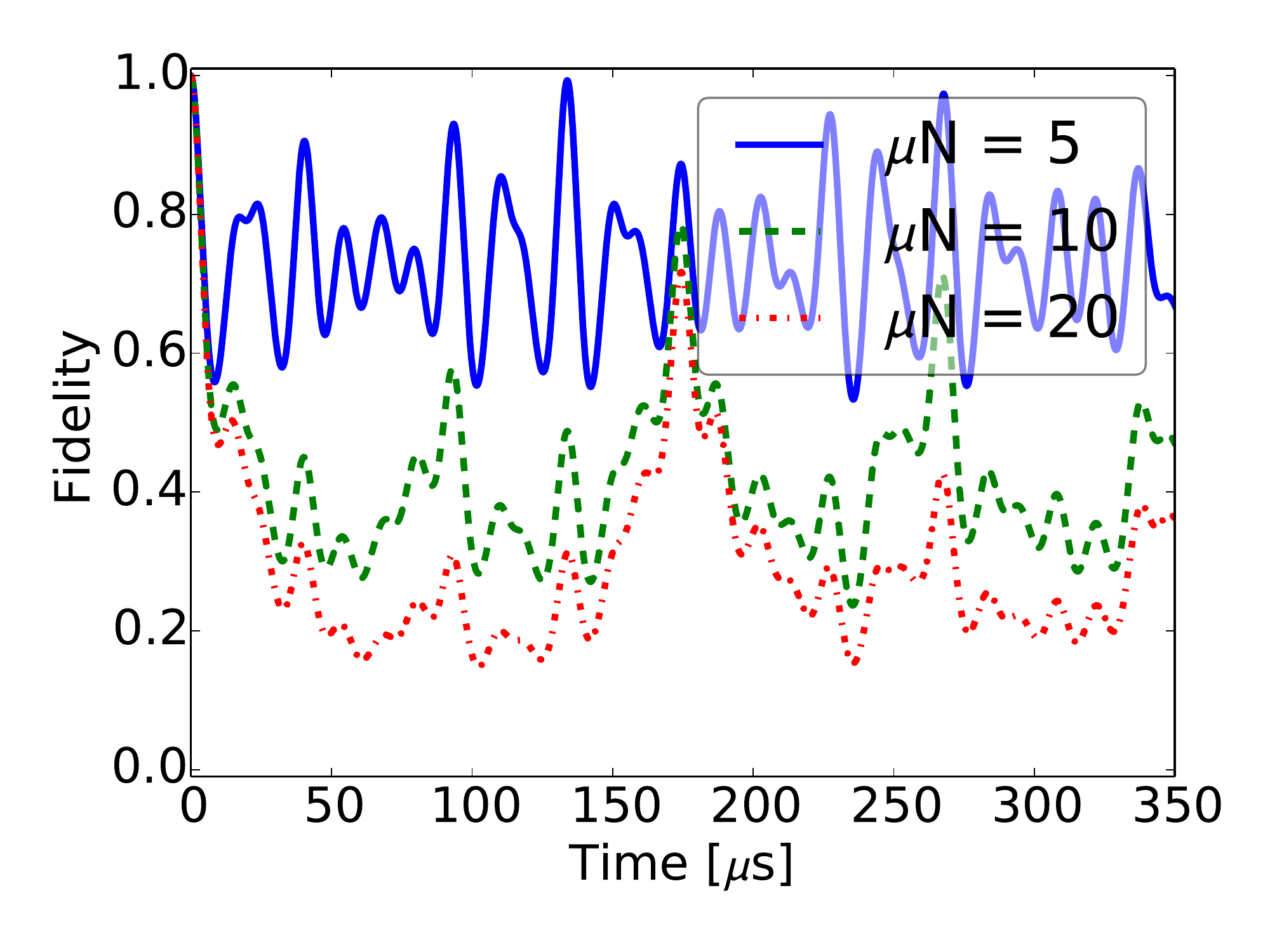}&
			\includegraphics[width=5.5cm]{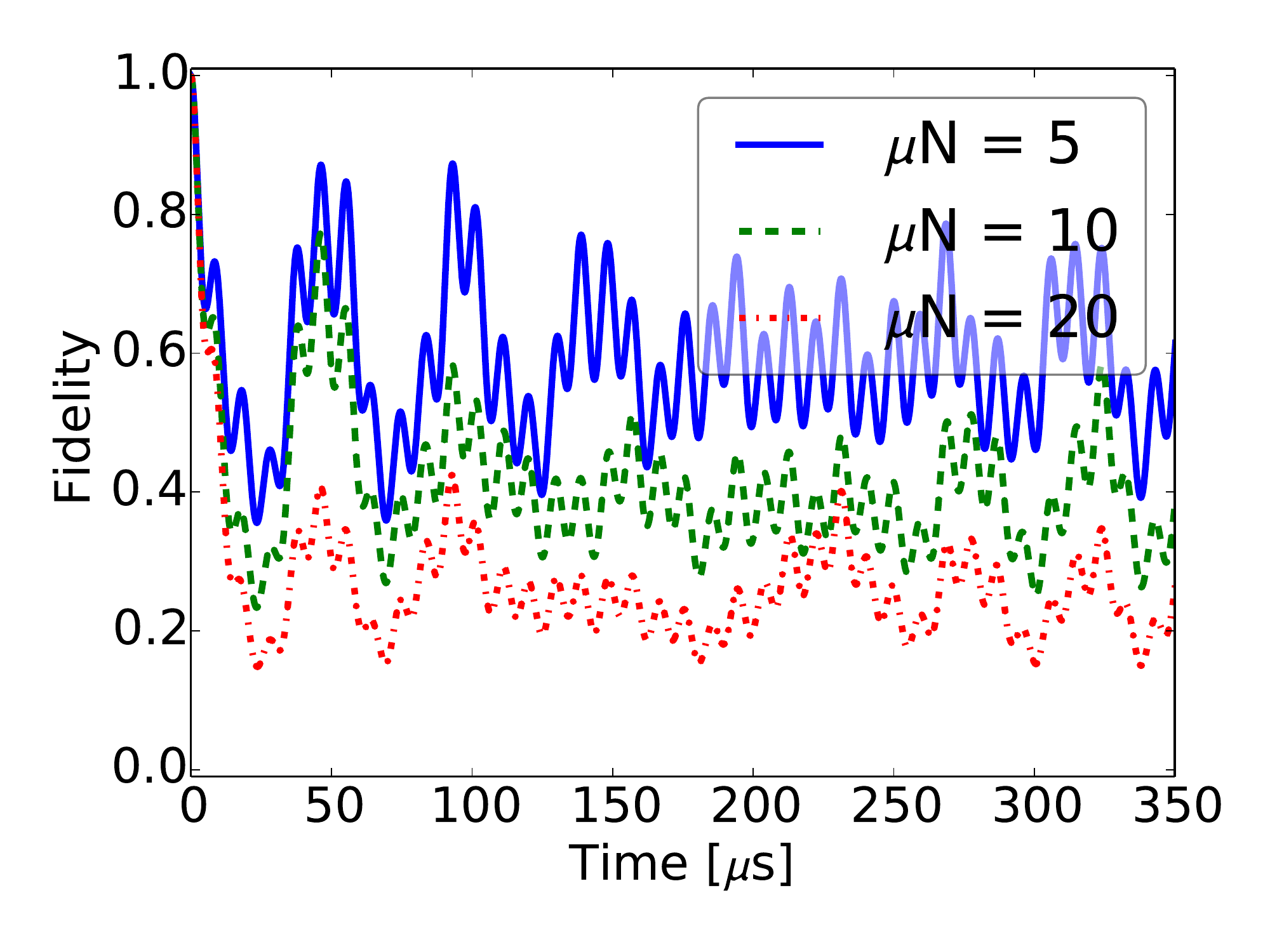}&
			\includegraphics[width=5.5cm]{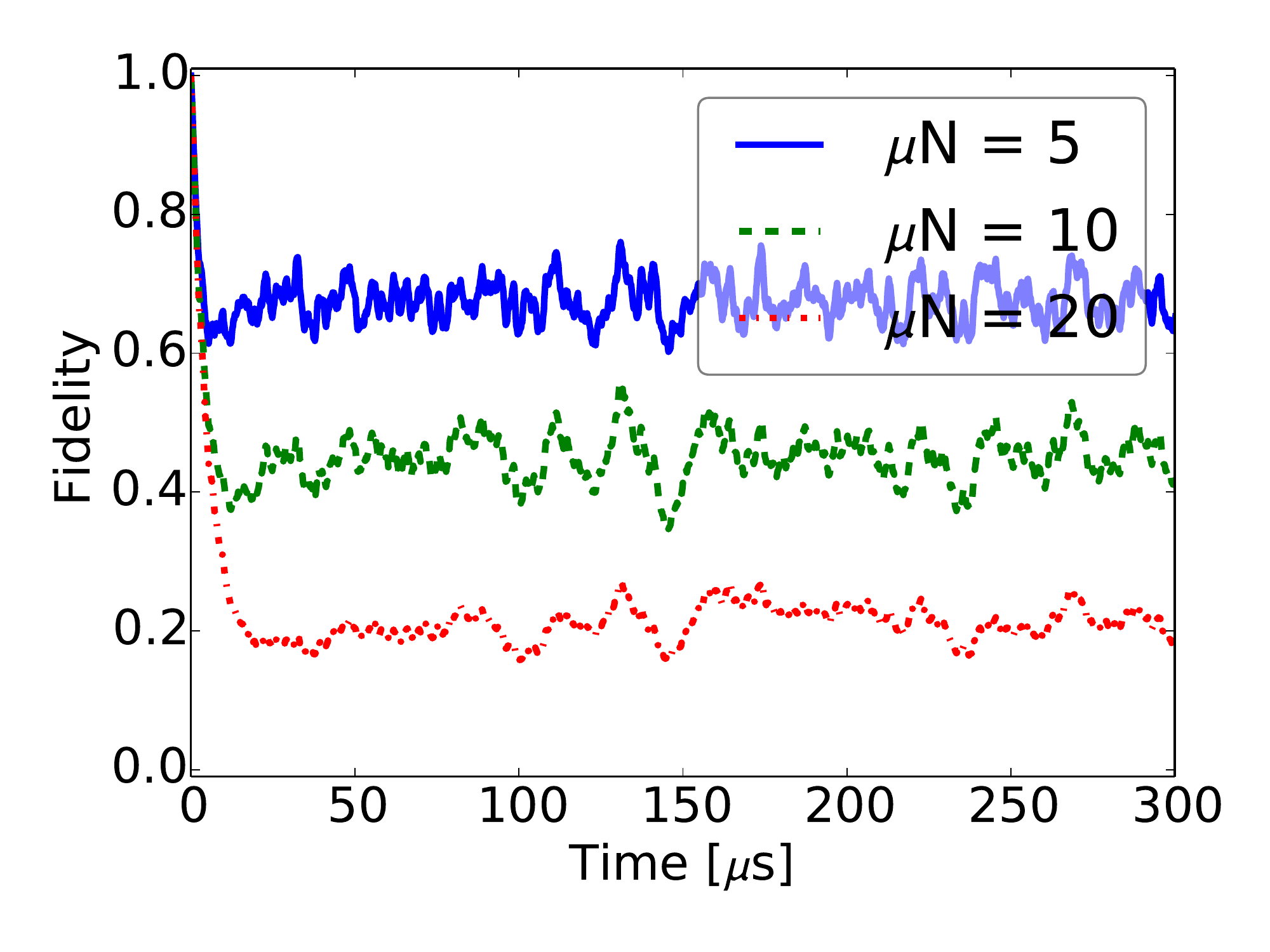}\\
			\textbf{(c)}&&\\
			&\includegraphics[width=5.5cm]{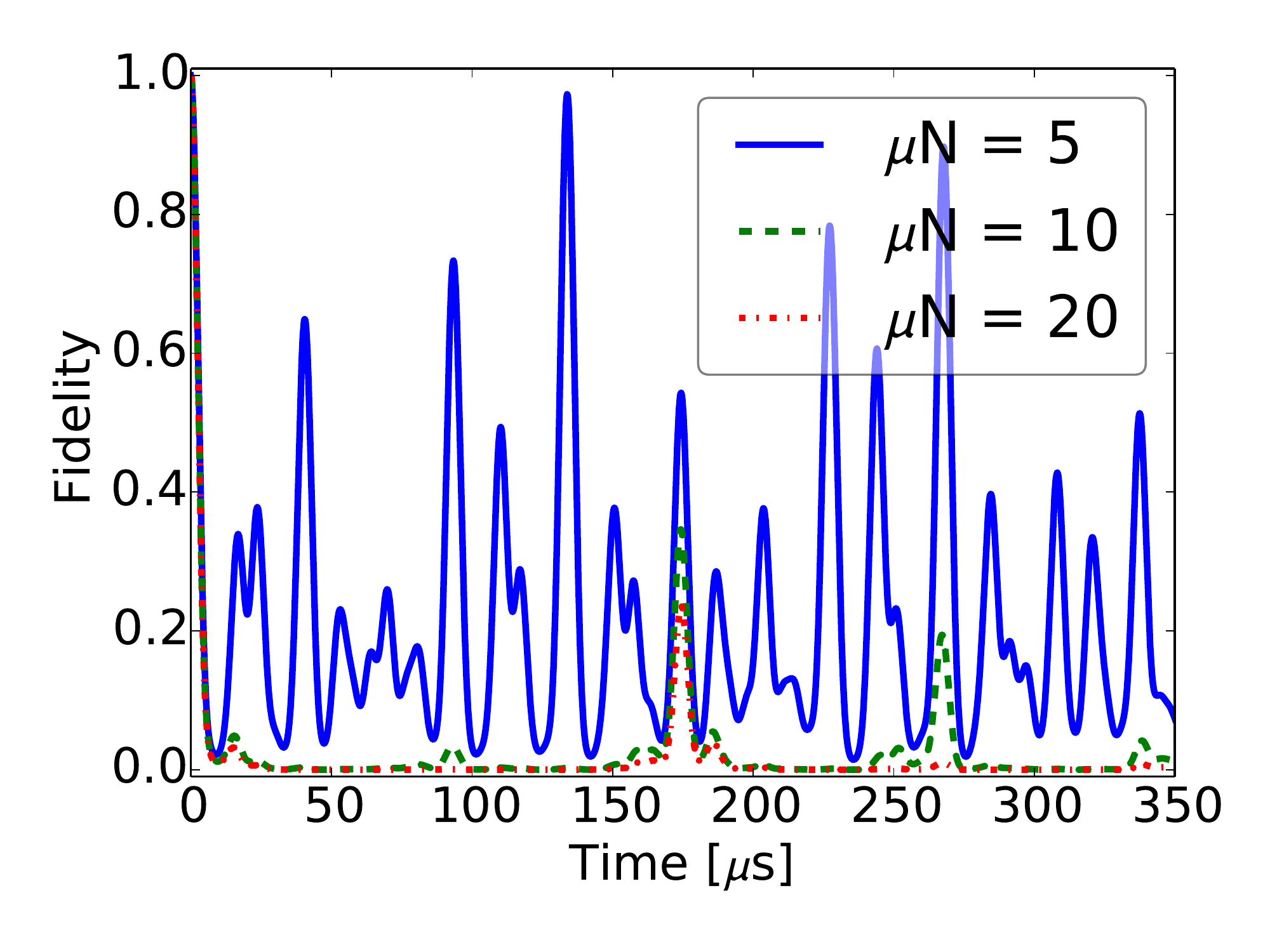}&
			\includegraphics[width=5.5cm]{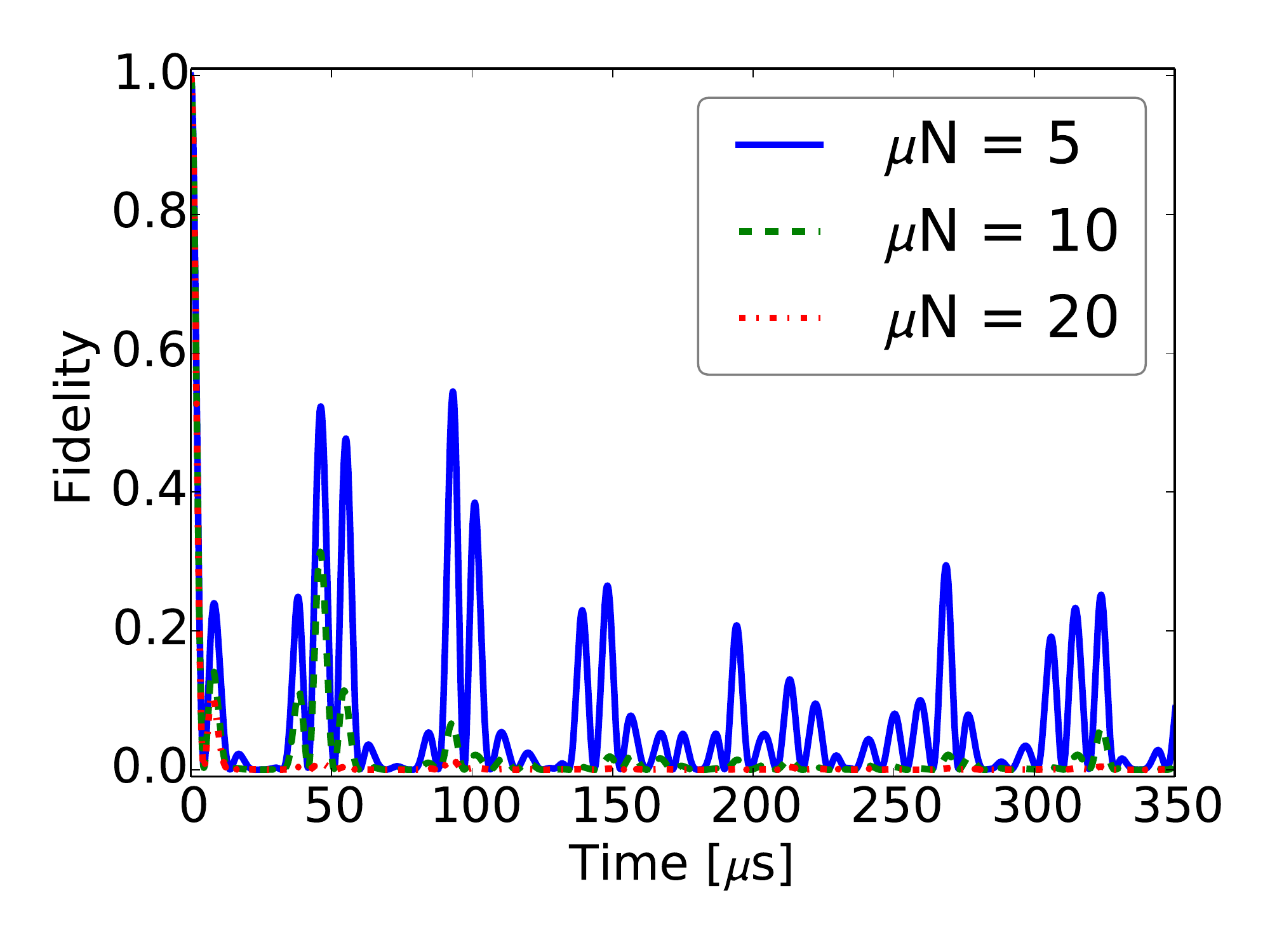}&
			\includegraphics[width=5.5cm]{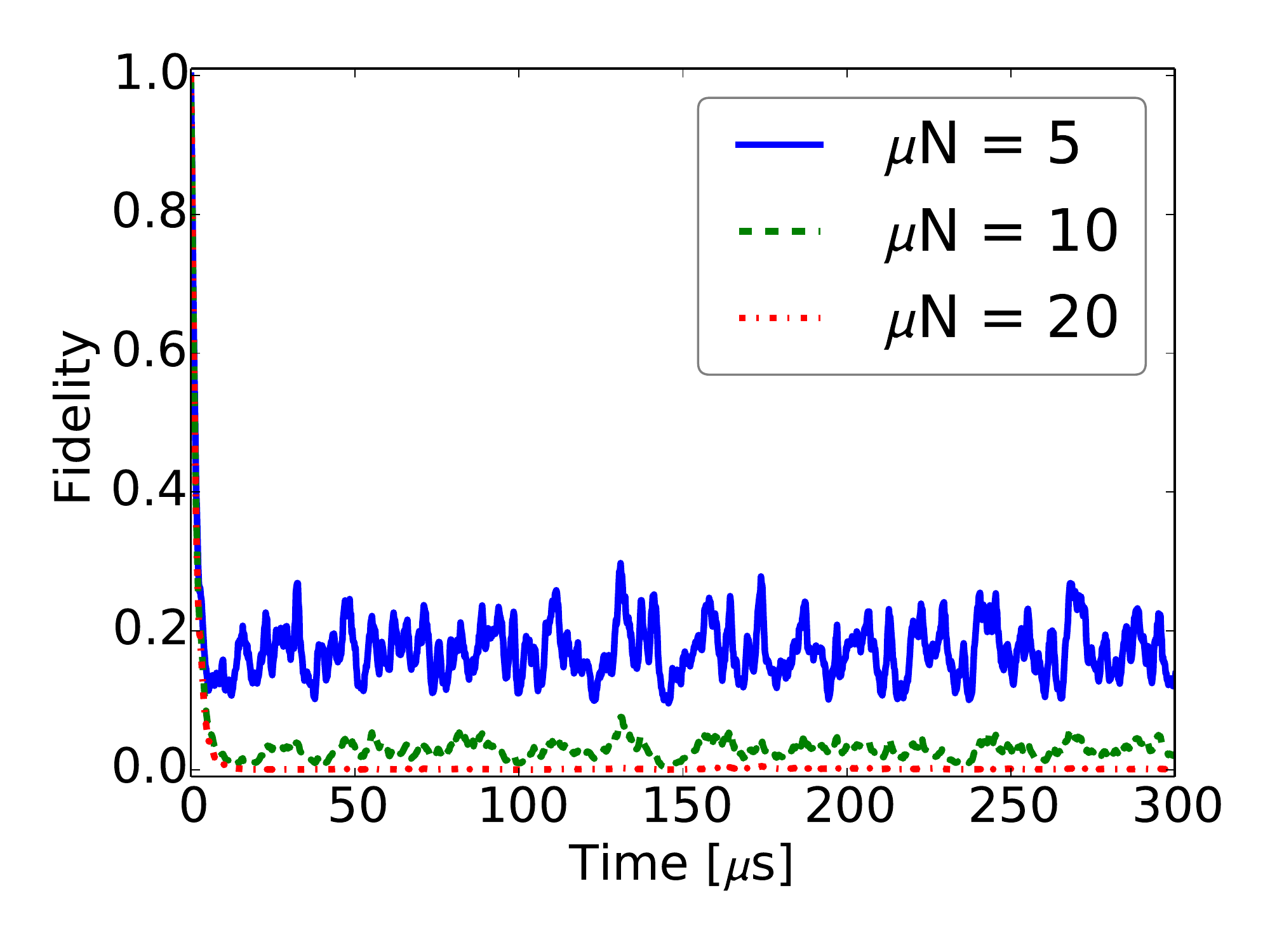}\\
		\end{tabular}
	\end{center}
	\caption{Fidelity as a function of evolution time for 2 spatial realizations, labelled (I) and (II), and an average over 100 realizations of the nuclear environment, corresponding to macrofractions containing $\mu N=$5, 10, 20 nuclear spins. Panels (a) -- (c) correspond to macrofraction polarizations $p$= 0.1, 0.5 and 1.0, respectively.}
	\label{fig:reals}
\end{figure*}

Analytical studies of the decoherence and the fidelity factors, derived in the previous Section, are quite limited due to the fact that compact approximate expressions can only be obtained for weakly coupled ($\omega \! \gg\! A^{z}_k$, $A^{\perp}_k$) or weakly polarized nuclei - and below we will show that having large polarization and strong coupling is needed for appearance of genuine SBS.
We will now present results of numerical investigations. We recall that in order to show the creation of SBS states, both functions (\ref{full_gamma}) and (\ref{Fmac}) must vanish. 

Experiments and theory of decoherence of NV centers show that the time-dependence of their dephasing is very prone to the effects connected with presence of a few, maybe few tens of strongly coupled nuclei, located 1-2 nm from the defect. A widespread collection of applications of such nuclei, either for sensing or creating a register for quantum networks have been discussed and tested experimentally \cite{Zhao_NN11,Zhao_NN12,Bradley2019,Abobeih2019}. Presence of such ``fingerprints'' of a spatial arrangement of environmental spins most strongly coupled to the qubit is also expected in the time-dependence of fidelity between the states of a macrofraction conditioned on two states of the qubit. Here we consider a given number $fN$ of nuclear spins within a ball of radius $r_p$ around NV center, which are in a polarized state. Outside of this region, the environment is initialized in a completely mixed state, which corresponds to room temperature conditions, typical for NV center experiments.

\begin{figure}[!htbp]
	\begin{center}
	\begin{tabular}{cc}
	(a)&\\
	&\includegraphics[width=0.83\linewidth]{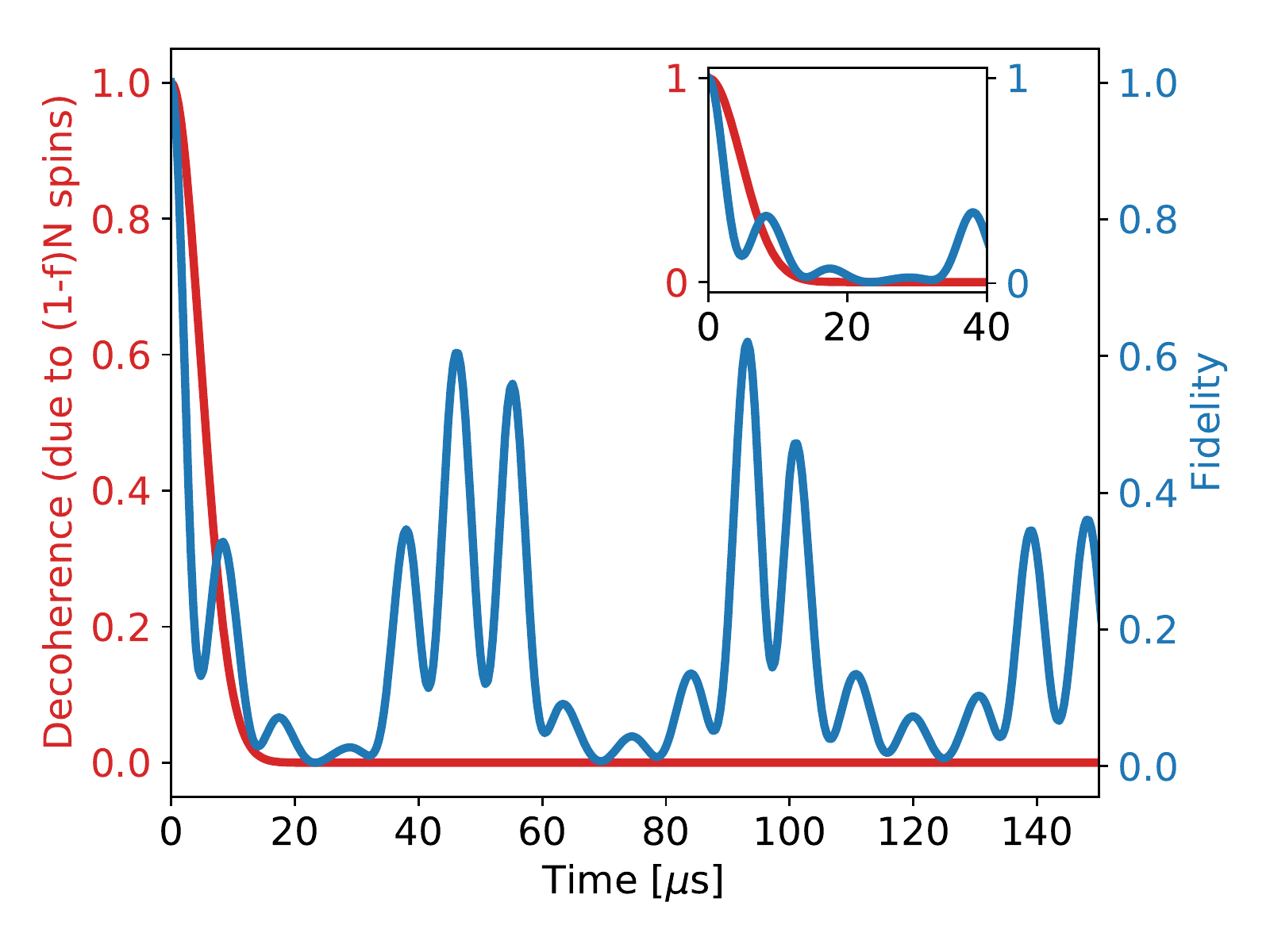}\\
	(b)&\\
	&\includegraphics[width=0.83\linewidth]{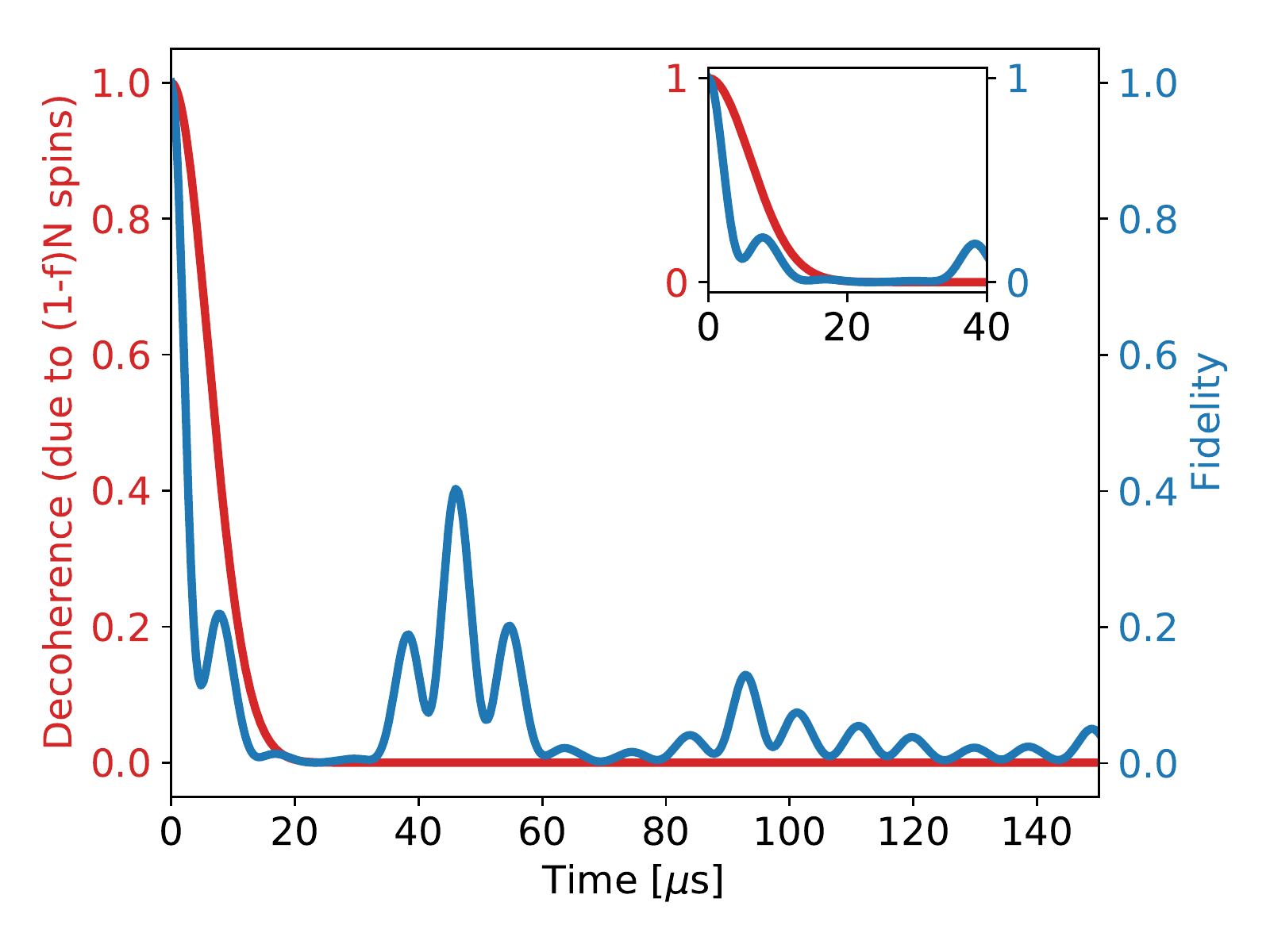}\\
	(c)&\\
	&\includegraphics[width=0.83\linewidth]{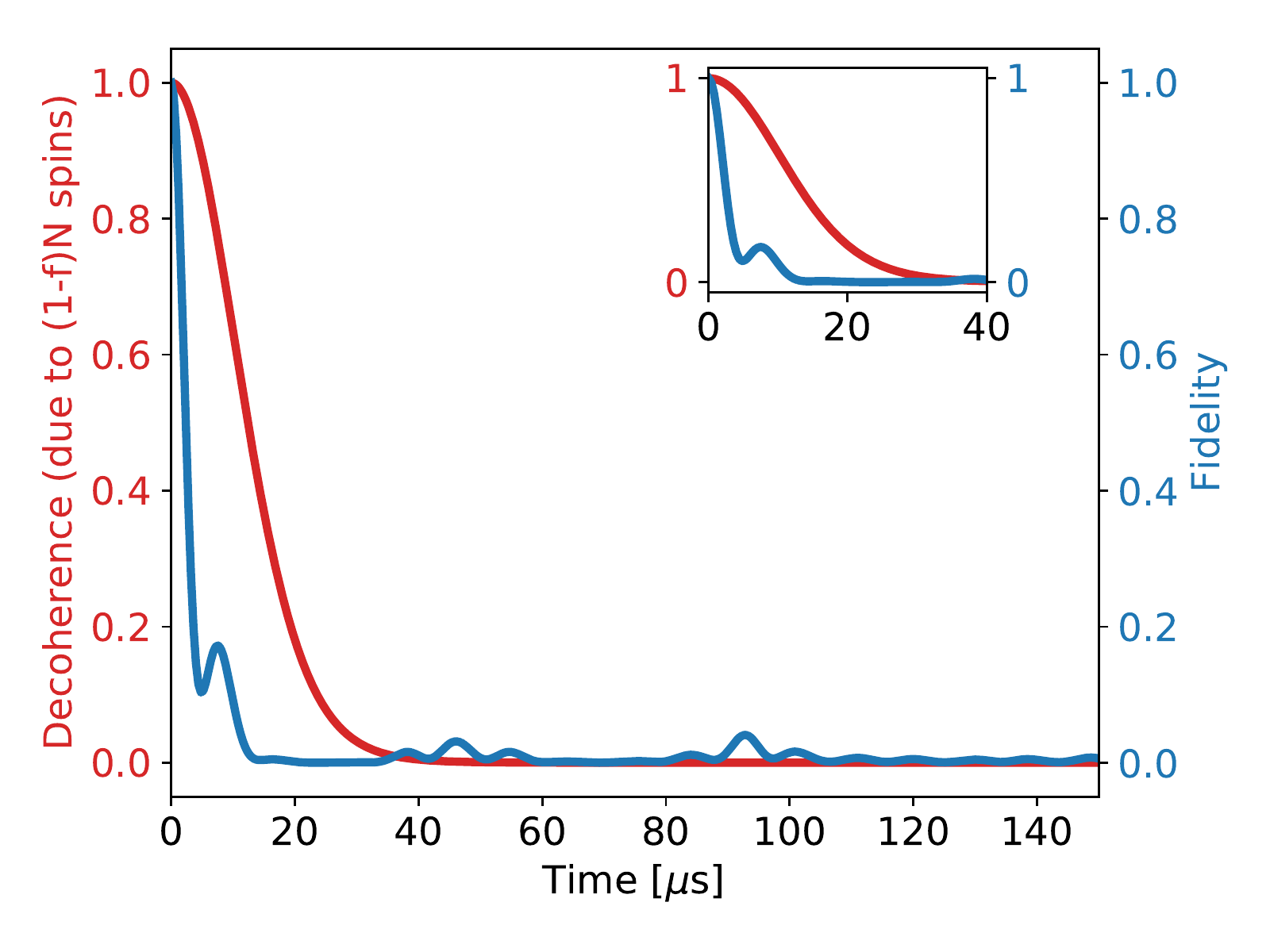}
	\end{tabular}
	\end{center}
	\label{fig:05}
	\caption{Example of SBS states formation. We assume realization (II) polarized up to $p=0.9$. The strongly coupled fraction $fN$ is divided into two identical fractions, $fN=2\times \mu N$,  of a size: (a) $\mu N$ = 5, (b) $\mu N$ = 10 and (c) $\mu N$ = 20. The blue curve corresponds to the fidelity and the red curve to decoherence due to the unobserved environment. Total number of spins in the nuclear bath is $N$ = 400. The insets show the short-time behavior that is close to Gaussian decay.}
\label{fig:SBS}
\end{figure}

	Numerical studies performed here are based on parameters of natural samples of diamonds implanted with nitrogen-vacancy centers. Diamond lattice symmetry corresponds to a diamond cubic crystal structure, with a cubic unit cell containing three tetrahedrons with carbon atoms as vertices. Each side of the unit cell corresponds to $a_{NN}=$0.357 nm distance between the neighboring carbon atoms. For a given realization of the environment around an NV center positioned at one of these vertices, positions of spinful $^{13}$C nuclei, described by lattice are drawn from a random uniform distribution of sets of three lattice indices, corresponding to spatial location of these species. Size of the environment, enumerated by number of spins in the environment - $N$ - 
	has to be estimated by the convergence of results for quantities of interest (decoherence factor due to unobserved fraction of the environment, fidelity between the conditional states for macrofractions) as function of size of part of environment taken into account, while considering these quantities on certain timescale (here determined by decoherence due to unobserved nuclei).
	 Experimental and theoretical works show that NV centers should be sensitive to nuclei at distances of a few nanometers with natural concentration of $^{13}$C isotope in the lattice \cite{Zhao2012,Dobrovitski_ARCMP13}. This corresponds to the total number of spins on the order of $N$=300 -- 500. We assume here $N=400$. The dipolar couplings of nuclei to the NV center, i.e. $A^{\perp}_k$ and $A^z_k$, are then determined from Eq.\eqref{eq:Ajk} and are random quantities due to random positions $\mathbf r_k$. We assume equal Zeeman splittings $\omega_k=\omega$, corresponding to application of constant external magnetic field $B=10$ Gauss. Concerning polarization degrees, we assume an experimentally viable scenario of application of DNP as a preparatory stage, which results in the environment split into highly polarized and non-polarized parts as depicted in Figure \ref{fig:pol} (a). We then associate the observed part of the environment $fE$, of the size $fN$, with the highly polarized fraction, assuming equal polarization for all spins in $fE$, $\forall_{k\in fE} \,\,p_k=p\neq 0$. The unobserved part $(1-f)E$, of the size $(1-f)N$, is then the unpolarized fraction, assumed initially in a completely mixed state: $\forall_{k\in (1-f)E}\,\,p_k=0$. Thus, the only randomness is in the coupling constants $A^{\perp}_k$ and $A^z_k$.
	
\begin{figure*}
	\begin{center}
		\begin{tabular}{cc}
			(I)&(II)\\
			\includegraphics[width=0.9\columnwidth]{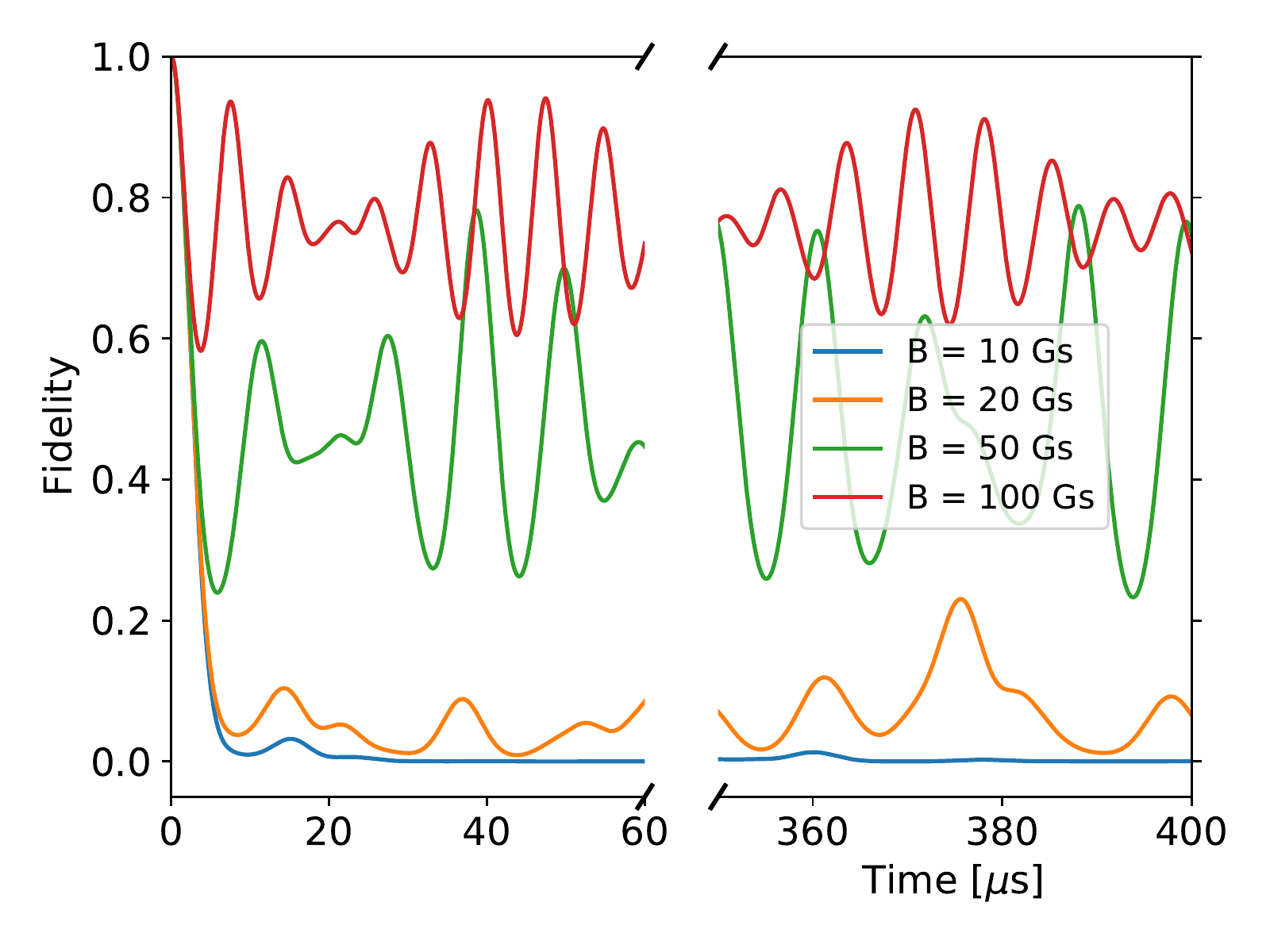}&
			\includegraphics[width=0.9\columnwidth]{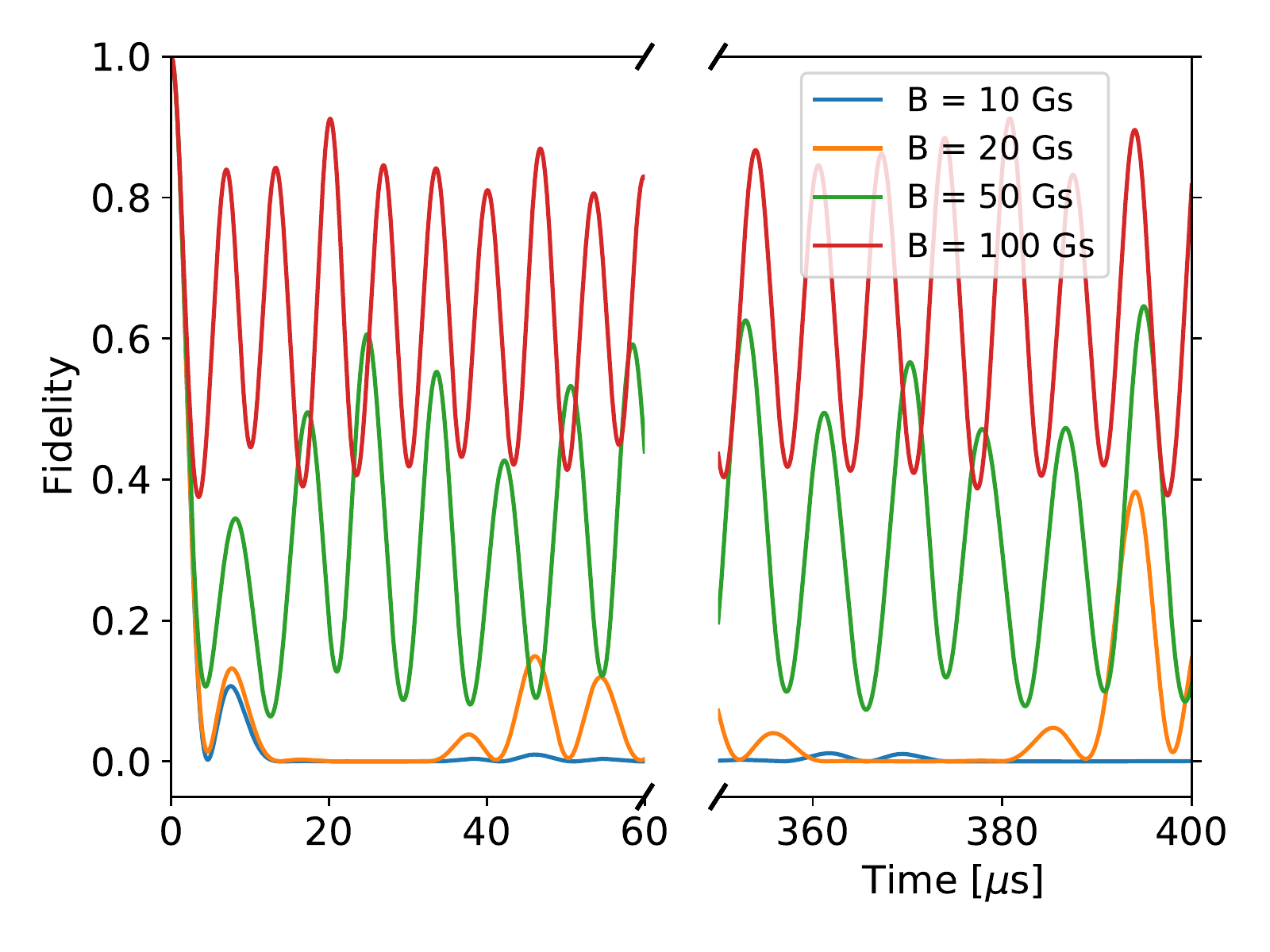}\\
		\end{tabular}
	\end{center}
	\caption{Fidelity for two spatial realizations of the nuclear environment.  We assume macrofractions of 20 completely polarized spins ($p$=1). Each curve corresponds to a different value of external magnetic field (in Gauss), as indicated in the legend. Panels (a) -- (b) represent the fidelity on the short and the long timescales respectively. }
	\label{fig:fields}
\end{figure*}
	

We will first look at the decoherence process as the necessary condition for the SBS formation. The choice of the unobserved environment by removing the strongly coupled nuclei from the decoherence function (\ref{full_gamma}) means that decoherence as a function of time can be well-represented by one spatial realization of the environment (assuming no application of any resonant operations on the central qubit). Figure \ref{fig:gamma} shows the squared modulus of the decoherence factor (\ref{full_gamma}) for a single sample realization, further denoted as  (II), and different fully polarized (observed) fractions $fN$. All of these curves have been tested for relevance of intra-environment interactions, using the so called Cluster-Correlation Expansion (CCE), using which one can account in a controllable way for influence of inter-nuclear interactions on qubit's decoherence \cite{Yang_CCE_PRB08,Yang_CCE_PRB09,Zhao2012,Yang_RPP17}. The calculations have shown that for magnetic field $B \! = \!10$G, it is sufficient to describe the decoherence dynamics due to the environmental remainder as non-interacting (CCE-1) on the timescales of $t<300 \mu$s, which is in agreement with \cite{Zhao2012,Liu2012}. Figure \ref{fig:gamma} shows a smooth Gaussian decay of coherences on the time scale of $10-20\mu$s, depending on how many of $N$ nuclei are left for observation.

 Let us now look at the state fidelity (\ref{Fmac}) for the polarized spins. The results are presented in Fig. \ref{fig:reals}. The first two columns  show (\ref{Fmac}) as a function of total evolution time for two different spatial realizations of the nuclear environment in relatively low magnetic field of 10 Gauss. As described earlier, nuclear spins are randomly uniformly distributed in the diamond lattice and their concentration is 1.1\%. The rows $(a)-(c)$ correspond to different polarizations, assumed the same for all the nuclei in the macrofracion. First of all, one can see that the polarization $p$ plays a crucial role in the fidelity behavior and for low polarization, Fig. \ref{fig:reals}a, there is no chance of approaching even remotely the state distinguishability (\ref{eq:perp}) for any reasonable macrofraction size. This is because for low $p$ the initial environment state is very close to the totally mixed stated (cf. (\ref{rho_init})) and hence is very little affected by the interaction. 
However, Figs. \ref{fig:reals}b,c show that for higher polarizations ($p>0.5$), even a macrofraction of few dozens nuclei can achieve some level of distinguishability
for times $t>100\mu s$ given our assumed parameters. These plots also show the initial Gaussian decay of the fidelity as predicted by the short-time analysis of Section \ref{sec:fidelity}. Past short times however, one can see an oscillatory behavior, especially prominent for small macrofraction sizes. This is due to the not enough randomization for small sizes $\mu N$ of the strongly coupled nuclei. 

In order to see more clearly the qualitative behavior of the fidelity, we present  its average over a hundred realizations of the positions of the $^{13}$C nuclei around the NV center in the last column of Fig. \ref{fig:reals}.  Clearly, the orthogonalization of the conditional environmental states is both faster, and more complete, for larger polarizations and macrofraction sizes. 

Comparing Fig. \ref{fig:gamma} and Fig. \ref{fig:reals}b,c suggests there is a time region when both functions come reasonable close to zero, indicating that the partially reduced state is close to the SBS form. Indeed it is so as Fig. \ref{fig:SBS} shows. Working with the realization $(II)$ from Fig. \ref{fig:reals} for definiteness, we assumed the polarized (observed) fraction $fE$ is divided into two identical macrofractions $fE = \mu E \cup \mu E$ of the size $\mu N$.
We assume the polarization degree $p=0.9$, which should be experimentally viable, close to achieving an initially pure state for the observed part of the environment. 
In the case of the division into $\mu N$=5 spin-macrofractions (cf. Fig. \ref{fig:SBS}a ), the fidelity strongly oscillates, indicating an insufficient number of spins in the macrofraction.  However, for $\mu N$=10, Fig. \ref{fig:SBS}b, although the fidelity shows some revivals for certain times, it generally tends to weakly oscillate around zero.
For $\mu N$=20, Fig. \ref{fig:SBS}c, the situation is even better with a definite decay of the fidelity past $t \! \approx \! 20\mu$s. Therefore, we can claim that for macrofractions of at least 10 strongly coupled nuclear spins in the highly polarized ($p\geq 0.9$) part of the environment, a SBS state is approached within 100 $\mu$s. Since in the current state of the art experiments, a polarization of around $20$ tightly coupled spins is achieved, the realistic SBS structure one can expect is a two-observer one.  
	
	All of the previous results have been calculated for a relatively low magnetic field of 10 Gauss. When increasing the field, we should be able to suppress the dynamics of nuclear spins induced by transverse the hyperfine couplings - i.e.~the dynamics that is caused by interaction with the qubit, and thus might be conditional on the state of the qubit.
	The dependence of the fidelity on the magnetic field is presented in Figure \ref{fig:fields} for a group of completely polarized 20 nuclear spins. For fields between 10-20 Gauss, the fidelity only slightly deviates from zero, but for 50-100 Gauss, it persistently oscillates, which means that only a few nuclear spins contribute to formation of mutually orthogonal conditional states and thus the formation of SBS is not observed. These few nuclear spins are the ones that are still {\it strongly coupled} to the qubit at elevated magnetic fields. By ``strong coupling'' we mean here that the characteristic energy scale of qubit-nucleus coupling (more precisely, of the part of the coupling that leads to qubit-dependent dynamics) is larger than the characteristic energy scale of Hamiltonian of the nucleus, i.e.~$A_{k}^{\perp} \! \gg \! \omega$. Only in this limit, in which the qubit-environment coupling $\hat{V}_m$ dominates over the environmental Hamiltonian, $\hat{H}_E$ (a condition known as the ``quantum measurement limit'' of decoherence, see \cite{Schlosshauer_book}), we can expect the qubit to leave a significant trace of its state (or even presence) on the state of the environment.

 When looking at statistics of hyperfine couplings for each of a hundred realizations of nuclear bath around NV center, as discussed in Appendix C, it becomes clear why $\mu N\approx 20$ corresponds to the formation of SBS: Around 15-20 nuclear spins closest to the NV center have the transverse hyperfine exceeding nuclear Zeeman splitting for $B=$10 Gauss. Additionally, for roughly a half of these spins also the component of the hyperfine coupling parallel to the magnetic field exceeds the Zeeman splittings. For these spins one cannot of course use the weak-coupling approximation, and one has to consider the full form of Eqs.~(\ref{eq:F01}) and (\ref{Fmac}) for fidelity. For a few strongly coupled nuclei, oscillations of fidelity with frequencies $\Omega_k \! \approx \! \sqrt{(A^\perp_k)^2 + (A^z_k)^2}$ should be indeed visible.

	\section{Conclusions}\label{sec:conclude}
We have analyzed a realistic model of NV center as a 'simulator' for an important process of the quantum-to-classical transition -- the appearance of objectivity. The latter is described by Spectrum Broadcast Structures -- specific multipartite quantum states, encoding an operational notion of objectivity and related to the idea of quantum Darwinism. From our theoretical analysis it follows that using current state of the art dynamical polarization technique, the post decoherence quantum state of the NV center and two macrofractions, each consisting  of about 10 strongly polarized nuclei localized close to the center, comes reasonably close to a SBS form, provided that we keep the external magnetic field below $\approx \! 20$ Gauss, so that the polarized nuclei close to the NV center are strongly coupled to it, i.e.~the energy scale of their coupling to the qubit exceeds their Zeeman energy. In these conditions, during the decoherence process the information about the state of the NV center qubit becomes redundantly encoded in its nearest environment in the strongest possible form, and hence becomes objective. 

This is, to the best of our knowledge, the first study of SBS using a model that closely describes a system that is actually a subject of ongoing experiments. 
Let us discuss the possibilities of an experimental verification of our results.

NV center is the only qubit in the considered system that can be directly read out. It is possible to create a coherent quantum state of nuclear spins or even an entangled state of NV center and a few nuclei \cite{Bradley_PRX19}, but then the tomography of such state is performed using NV center coherence.
Therefore, a direct observation of an SBS state or measurement of fidelity between conditional states of observed fraction is not possible in a setup with a single NV center qubit, since the state comes into being as a result of central qubit decoherence. However, according to \cite{Bradley_PRX19}, tomography of conditional states of the bath or at least identification of timescales for orthogonalization of conditional states of the observed bath as discussed in this work should be experimentally viable. One could take advantage of the ability of the NV center qubit to characterize nearby nuclear spins in a two-qubit setup, in which the second qubit is kept in $m\!= \!0 $ state (decoupled from the environment) while the first one decoheres, and only after time at which creation of SBS is expected, it is rotated into a superposition state, and its dephasing under dynamical decoupling is used to characterize the state of the nuclear environment common to the two qubits. In order for polarized spins close to the first qubit to be within such a common environment, the distance between the centers should be a few nanometers \cite{Kwiatkowski_PRB18}, which will be challenging to achieve, but it's not inconceivable, with entanglement of two centers separated by $\approx \! 20$ nm achieved a few years ago \cite{Dolde_NP13}.

	\section*{Acknowledgements}
   We acknowledge the financial support by Polish National Science Center (NCN), Grant no.~2015/19/B/ST3/03152 (\L{}C), Grant no.~2019/35/B/ST2/01896 (JKK), and PhD Student Scholarship no.~2018/28/T/ST3/00390 (DK).


		\appendix


	\section{Decoherence of a qubit defined between states $m$ and $m'$ due to free evolution with a non-interacting bath}
	Coherence of a qubit defined between $m$ and $m'$ states evolving freely with a bath of non-interacting spins (from the unobserved part of the environment), in a rotating frame with respect to free Hamiltonian of the qubit, can be expressed as:
	\begin{equation}
	\gamma_{mm'}(t)=\prod\limits_{k\in (1-\mu)E}\gamma_{mm'}^k(t),
	\end{equation}
	with the contribution from a single spin $k$:

	\begin{widetext}
		\begin{align}
		&\gamma_{mm'}^k(t)=\cos\frac{\omega_m t}{2}\cos\frac{\omega_{m'}t}{2}+\frac{mm'(A^{\perp}_k)^2+(A^z_km+\omega_k)(A^z_k m'+\omega_k)}{\omega_m\omega_{m'}}\sin\frac{\omega_m t}{2}\sin\frac{\omega_{m'}t}{2}-\nonumber\\
		&-ip_k\left(\frac{A^z_k m+\omega_k}{\omega_m}\cos\frac{\omega_{m'}t}{2}\sin\frac{\omega_mt}{2}-\frac{A^z_k m'+\omega_k}{\omega_{m'}}\cos\frac{\omega_m t}{2}\sin\frac{\omega_{m'}t}{2}\right)
		\end{align}
	\end{widetext}

		\begin{figure*}
	\begin{center}
		\includegraphics[width=0.77\textwidth]{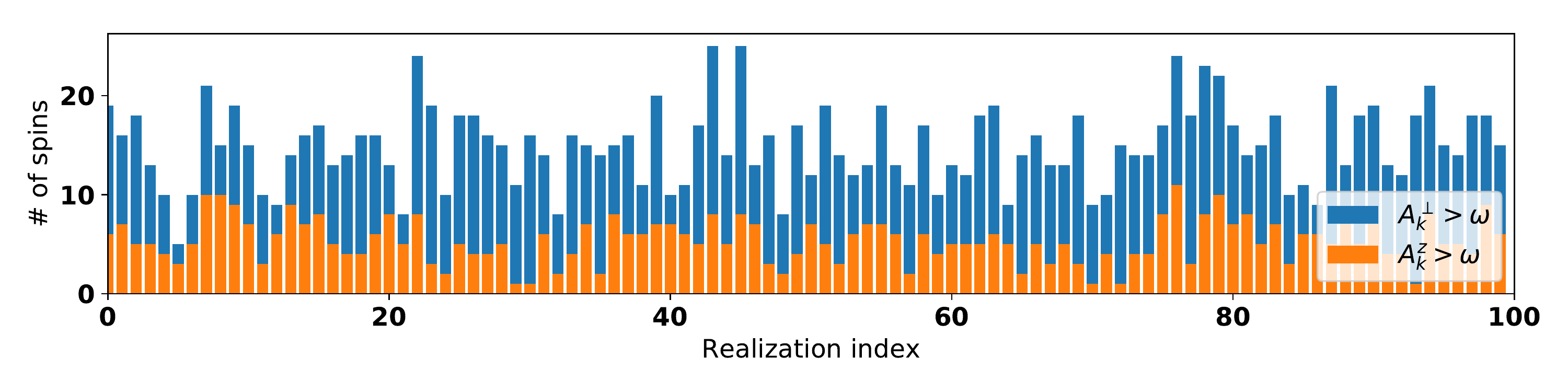}
	\end{center}
	\caption{Number of nuclear spins for which transverse or parallel hyperfine interaction value is larger than nuclear Zeeman splitting for $B=10$ Gauss.}
	\label{fig:stats}
\end{figure*}
	\section{Fidelity between environmental states conditioned on qubit defined between states $m$ and $m'$}

	For a qubit defined between arbitrary $m$ and $m'$ states, fidelity for noninteracting bath can be expressed as stated in Eq. (\eqref{eq:generalFidelity}):
	\begin{equation}
	\F_{mm'}^{\mu E}(t)\equiv \F(\hat{\rho}^{
		\mu E}_{m}(t),\hat{\rho}^{\mu E}_{m'}(t))=\prod\limits_{k\in\mu N}\F(\hat{\rho}^{
		k}_{m}(t),\hat{\rho}^{k}_{m'}(t)).
	\end{equation}
	Contribution for a single member of such macrofraction corresponds to the following formula:
	\begin{widetext}
		\begin{align}
		&\F(\hat{\rho}^{
			k}_{m}(t),\hat{\rho}^{k}_{m'}(t))=1+p_k^2\left[(A^{\perp}_k)^2\left(\frac{m^2}{\omega_{m
			}^2}\sin^2\frac{\omega_{m}t}{2}-\frac{m'^2}{\omega_{m'
			}^2}\sin^2\frac{\omega_{m'}t}{2}\right)+\right.\nonumber\\
		&+2 mm'\left(\frac{(A^{\perp}_k)^2(m A^z_k+\omega_k)(m'A^z_{k}+\omega_k)}{\omega_m^2\omega_{m'}^2}\sin\frac{\omega_m t}{2}\sin\frac{\omega_{m'}t}{2}+\frac{(A^{\perp}_k)^2}{\omega_m\omega_{m'}}\sin\omega_m t\sin\omega_{m'}t\right)\nonumber\\
		&\left.+\frac{2m^2m'^2(A^{\perp}_{k})^4}{\omega_m^2\omega_{m'}^2}\sin^2\frac{\omega_m t}{2}\sin^2\frac{\omega_{m'}t}{2}\right]
		\end{align}
	\end{widetext}

	From the form of this expression one can observe that Eq. \eqref{Fmac}, which corresponds to the case when qubit defined between $m=0$ and $m'=1$, simply reduces the above equation to one term proportional to $m^2=1$. For a qubit defined between $m=-1$ and $m'=1$, one needs to consider a complete expression.

\section{Fidelity for very strongly coupled nuclear spins}

	In the limit of very strong coupling to the qubit, i.e. when $A^{z}_k\gg \omega_k$ and $t\ll\frac{1}{\omega_k}\sqrt{1+\frac{(A^{\perp}_k)^2}{(A^{z}_k)^2}}$, this formula becomes:
	\begin{align}
	&\F^{\mu E}(t)\approx\nonumber\\
	&\approx\prod\limits_{k \in \mu N}\left[1-p_k^2\frac{(A^{\perp}_k)^2}{|A_k|^2}\left(1-\frac{2\omega A^z_k}{|A_k|^2}\right)\sin^2\left(\frac{t}{2}|A_k|\right)\right],
	\end{align}
	where $|A_k|=\sqrt{(A_k^\perp)^2+(A_k^z)^2}$.
	This limit can either correspond to strong oscillations observable on the timescale of orthogonalization of qubit-conditional states of a given macrofraction or, when exceeding a certain number of such spins, a rapid decay of fidelity as a function of total evolution time.

When looking at statistics of hyperfine couplings for each of a hundred realizations of nuclear bath around NV center, as represented in the Figure \ref{fig:stats} it becomes clear why $N\approx 20$ corresponds to formation of SBS, as around 15-20 nuclear spins closest to the NV center should have transverse hyperfine couplings which exceed nuclear Zeeman splitting for $B=$10 Gauss. Additionally, for roughly a half of these spins also component of the hyperfine coupling parallel to the magnetic field exceeds the Zeeman splittings. For these spins it is not practical to discuss the relevance of weakly coupled bath, however one should expect that the oscillations observed in fidelity for high magnetic fields, corresponds to dynamics of a few nuclear spins.

\newpage
%

\end{document}